\documentclass[twocolumn]{aastex701}  

\usepackage{xcolor}
\usepackage{textgreek}
\usepackage[utf8]{inputenc}
\usepackage[english]{babel}

\usepackage{hyperref}
\hypersetup{
    unicode, 
    colorlinks=true,
    linkcolor=linkcolor,
    citecolor=linkcolor,
    filecolor=linkcolor,
    urlcolor=linkcolor,
}
\usepackage{color,colortbl}
\definecolor{linkcolor}{rgb}{0.0,0.3,0.5}
\usepackage{tensind}
\tensordelimiter{?}
\DeclareGraphicsExtensions{.bmp,.png,.jpg,.pdf}
\usepackage{verbatim}
\usepackage[normalem]{ulem}
\usepackage{orcidlink}
\usepackage{soul}
\usepackage{xspace}

\newcommand{\Sec}[1]{{\protect\hyperref[sec:#1]{Section~\ref*{sec:#1}}}}

\newcommand{\Fig}[1]{{\protect\hyperref[fig:#1]{Figure~\ref*{fig:#1}}}}

\newcommand{\Figto}[2]{{\protect\hyperref[fig:#1]{Figures~\ref*{fig:#1}}~-~\ref{fig:#2}}}
\newcommand{\subFig}[2]{{\protect\hyperref[fig:#1]{Figure~\ref*{fig:#1}~#2}}}
\newcommand{\Equ}[1]{{\protect\hyperref[equ:#1]{Equation~\ref*{equ:#1}}}}

\newcommand{\Tab}[1]{{\protect\hyperref[tab:#1]{Table~\ref*{tab:#1}}}}

\newcommand{\App}[1]{{\protect\hyperref[app:#1]{Appendix~\ref*{app:#1}}}}

\newcommand{\numps}{1,355\xspace}
\graphicspath{ {./} }

\begin{document}
\title{Pixelated Posterior Sampling of Point Spread Functions in Astronomical Images}

\author[orcid=0000-0002-9086-6398]{Connor Stone}
\affiliation{Ciela - Montreal Institute for Astrophysical Data Analysis and Machine Learning, Montr{\'e}al, Canada}
\affiliation{Universit{\'e} de Montr{\'e}al, Montr{\'e}al, Canada}
\affiliation{Mila - Quebec Artificial Intelligence Institute, Montr{\'e}al, Canada}
\affiliation{University of Toronto, Toronto, Canada}
\email{connorstone628@gmail.com}

\author[orcid=0000-0001-9459-6316]{Ronan Legin}
\affiliation{Ciela - Montreal Institute for Astrophysical Data Analysis and Machine Learning, Montr{\'e}al, Canada}
\affiliation{Universit{\'e} de Montr{\'e}al, Montr{\'e}al, Canada}
\affiliation{Mila - Quebec Artificial Intelligence Institute, Montr{\'e}al, Canada}
\email{}

\author[orcid=0000-0001-8806-7936]{Alexandre Adam}
\affiliation{Ciela - Montreal Institute for Astrophysical Data Analysis and Machine Learning, Montr{\'e}al, Canada}
\affiliation{Universit{\'e} de Montr{\'e}al, Montr{\'e}al, Canada}
\affiliation{Mila - Quebec Artificial Intelligence Institute, Montr{\'e}al, Canada}
\email{}

\author[orcid=0009-0009-5797-0300,gname=Esmeralda,sname='Whitammer']{Esmeralda Whitammer}
\affiliation{University of Edinburgh, Edinburgh, Scotland}
\affiliation{Universit{\'e} de Montr{\'e}al, Montr{\'e}al, Canada}
\affiliation{Mila - Quebec Artificial Intelligence Institute, Montr{\'e}al, Canada}
\email{}

\author[orcid=0009-0008-5839-5937]{Gabriel Missael Barco}
\affiliation{Ciela - Montreal Institute for Astrophysical Data Analysis and Machine Learning, Montr{\'e}al, Canada}
\affiliation{Universit{\'e} de Montr{\'e}al, Montr{\'e}al, Canada}
\affiliation{Mila - Quebec Artificial Intelligence Institute, Montr{\'e}al, Canada}
\email{}

\author[orcid=0000-0003-3544-3939]{Laurence Perreault-Levasseur}
\affiliation{Ciela - Montreal Institute for Astrophysical Data Analysis and Machine Learning, Montr{\'e}al, Canada}
\affiliation{Universit{\'e} de Montr{\'e}al, Montr{\'e}al, Canada}
\affiliation{Mila - Quebec Artificial Intelligence Institute, Montr{\'e}al, Canada}
\affiliation{Center for Computational Astrophysics, Flatiron Institute, 162 5th Avenue, 10010, New York, NY, USA}
\affiliation{Trottier Space Institute, Montr{\'e}al, Canada}
\affiliation{Perimeter Institute, Waterloo, Canada}
\email{}

\author[orcid=0000-0002-8669-5733]{Yashar Hezaveh}
\affiliation{Ciela - Montreal Institute for Astrophysical Data Analysis and Machine Learning, Montr{\'e}al, Canada}
\affiliation{Universit{\'e} de Montr{\'e}al, Montr{\'e}al, Canada}
\affiliation{Mila - Quebec Artificial Intelligence Institute, Montr{\'e}al, Canada}
\affiliation{Center for Computational Astrophysics, Flatiron Institute, 162 5th Avenue, 10010, New York, NY, USA}
\affiliation{Trottier Space Institute, Montr{\'e}al, Canada}
\affiliation{Perimeter Institute, Waterloo, Canada}
\email{}

\begin{abstract}
    We introduce a novel framework for upsampled Point Spread Function (PSF) modeling using pixel-level Bayesian inference. 
    Accurate PSF characterization is critical for precision measurements in many fields including: weak lensing, astrometry, and photometry. 
    Our method defines the posterior distribution of the pixelized PSF model through the combination of an analytic Gaussian likelihood and a highly expressive generative diffusion model prior, trained on a library of HST ePSF templates. 
    Compared to traditional methods (parametric Moffat, ePSF template-based, and regularized likelihood), we demonstrate that our PSF models achieve orders of magnitude higher likelihood and residuals consistent with noise, all while remaining visually realistic. 
    Further, the method applies even for faint and heavily masked point sources, merely producing a broader posterior.
    By recovering a realistic, pixel-level posterior distribution, our technique enables the first meaningful propagation of detailed PSF morphological uncertainty in downstream analysis.
    An implementation of our posterior sampling procedure is available on GitHub.
\end{abstract}


\section{Introduction}
\label{sec:intro}

All astronomical observations are subject to image degradation arising from the redistribution of true sky flux due to the instrument's physical characteristics. 
Distortions originating from imperfect optics, telescope jitter, scattered light, atmospheric turbulence (for ground based observatories), and diffraction are inescapable realities of real world observation~\citep{Liaudat2023}.
The composite effect of these distortions is formalized as the Point Spread Function (PSF): the observed image of an ideal point-like light source. 
Accurate characterization of the PSF remains a persistent challenge in observational astronomy~\citep{Krist1993, Gunn1998, Perrin2012, Abraham2014, Liaudat2023, Schutt2025, Cuillandre2025}. 
Throughout this work, we refer to the effective PSF (ePSF), which encompasses not only the optical path PSF but also detector-level effects such as pixel response and charge diffusion. 
Critically, the PSF is not static. 
For space-based observatories like the Hubble Space Telescope (HST), the ePSF exhibits significant spatiotemporal variability driven by mechanical and thermal variations over time, such as orbital focus changes or ``breathing''. 
This variability, which also depends on position in the detector and wavelength, means that every observation is convolved with an essentially unique PSF, necessitating observation-specific modeling.

The accurate characterization of the PSF, and robust quantification of its morphological uncertainty, is critical for a broad spectrum of astronomical investigations. 
For example, precision shape measurements for weak gravitational lensing surveys are fundamentally limited by knowledge of the PSF~\citep{Farrens2022, Liaudat2023}. 
An accurate and well-defined PSF model underpins the structural parameters derived from galaxy scaling relations~\citep{vanderWel2014} and is essential for identifying fine-scale structures, such as the effect of dark matter subhalos in strong gravitational lensing analyses~\citep{Vegetti2010}. 
Beyond structural analysis, the PSF is foundational for high-precision astrometry~\citep{Anderson2000, Bennet2024} and accurate flux measurements~\citep{Stetson1987, Kuijken2008}. 
In high-contrast imaging, a precise PSF model is required for robust subtraction of stellar flux to reveal faint companions~\citep{Galicher2024}, while Active Galactic Nuclei (AGN) studies rely on it to deblend the central point source from its host galaxy~\citep{Zhuang2024}. 
In all these applications, an accurate and precise PSF model is crucial to avoid bias in even fundamental measurements such as photometry; further the PSF model uncertainty is a source of systematic error if not accounted for correctly. 

Given the recognized importance of the PSF, a diverse array of methodologies have been developed for its determination. 
These can be broadly separated into synthetic and empirical approaches. 
Synthetic methods utilize a first-principles approach, such as the analytic Airy disk pattern produced by a circular aperture~\citep{cambridge1864, herschel1899}.
More sophisticated wavefront-propagation approaches compute the expected PSF from detailed instrument models~\citep{Perrin2012, Desdoigts2023}. 
These may capture faint details that are otherwise challenging to extract, and can connect the components of the PSF with physical characteristics of the telescope.
However, the complex, time-dependent nature of the PSF often necessitates its direct empirical inference from observed images.

The simplest empirical strategy involves analyzing isolated stars. 
This has traditionally been accomplished through parametric modeling, which fits an analytic function (e.g., a Moffat profile) to the star image~\citep{Moffat1969, Peng2002, Stone2023}. 
While computationally efficient, this approach is realistically unable to capture the complex, high-frequency structures of a diffraction-limited PSF. 
More flexible basis methods, using Principal Component Analysis (PCA)~\citep{Jee2007}, shapelets\citep{Refregier2003, Paulin2008}, or starlets~\citep{Michalewicz2023}, can trade off between detail and computational complexity. 
Other methods rely on stacking many stars to build a high-S/N ePSF template~\citep{Anderson2006}, but this yields a fixed point estimate that does not perfectly represent the unique PSF of a given observation.

Fundamentally, pixel-level forward modeling represents the most principled, data-driven methodology for accurate PSF recovery.
This approach models the PSF on a high-resolution pixel grid, offering the flexibility to capture arbitrary morphological features~\citep{Bertin2011, Jarvis2021, Liu2022}.
An upsampled pixel-level PSF model will have more free parameters than observations to constrain them, making this a severely ill-posed problem.
Traditionally, this is overcome via regularization~\citep{Bertin2011}, or an implicit prior learned by a machine learning model~\citep{Herbel2018, Stark2018}.

A Bayesian framework turns the ill-posed problem of recovering the upsampled PSF given low resolution noisy samples into the well-posed problem of determining the posterior distribution $P(M| D)$ over PSF models ($M$) given some data ($D$).
Via Bayes' Theorem, this posterior is defined by a prior over PSF models $P(M)$ and a likelihood for the observed data $P(D|M)$.
The likelihood may be formulated exactly under the assumption of independent Poisson-distributed pixel noise (often approximated as a Gaussian in the high count limit).
The core challenge in Bayesian PSF inference is the specification of a realistic prior, $P(M)$. 
The complexity of PSF structures forced previous works to rely on simple, ad-hoc approximations for this prior.
In fact, all traditional methods described above may be understood as priors imposed to restrict the solution space; though they are highly miss-specified in this context.

\begin{figure*}
    \centering
    \includegraphics[width=\linewidth]{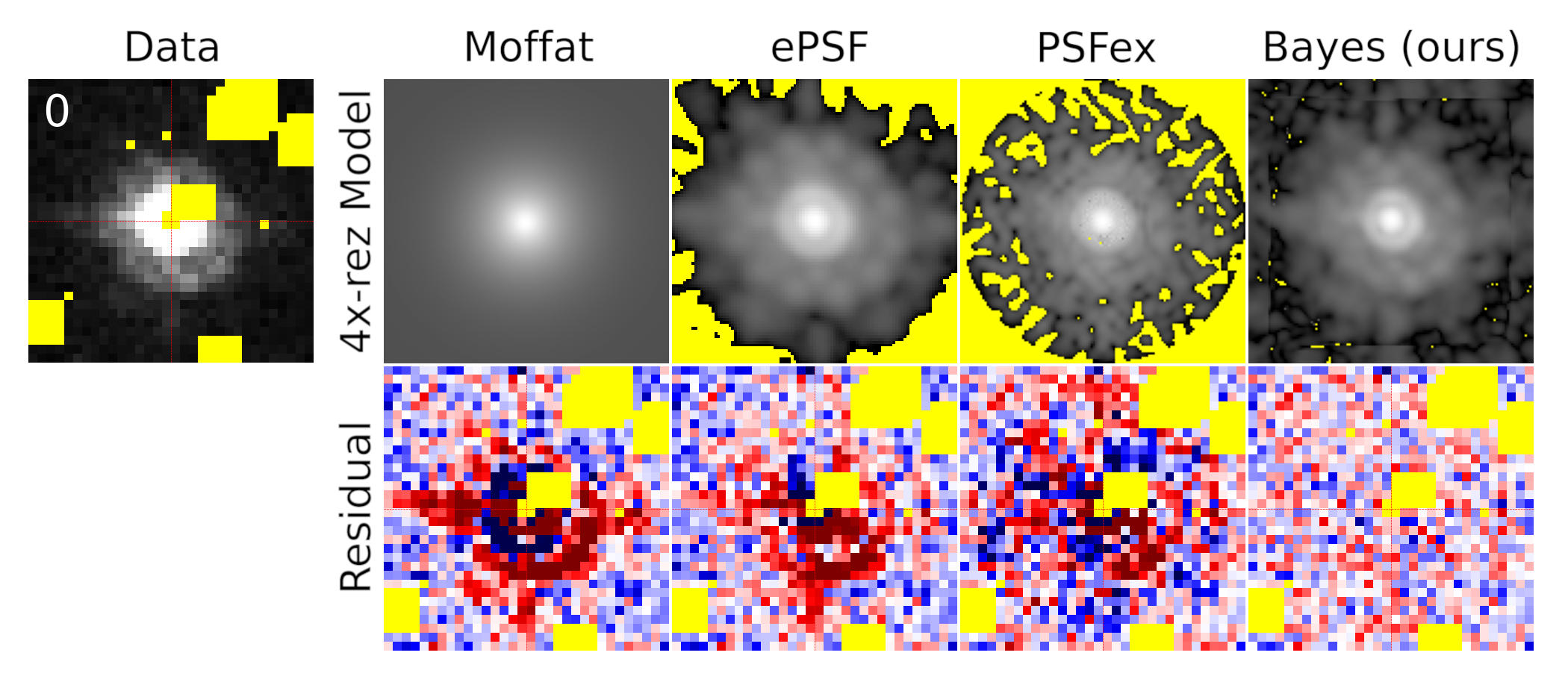}
    \caption{Methodological comparison. For a bright point source data cutout, we present three traditional modeling results alongside a single posterior sample from our procedure. The Moffat model includes only a smooth decline in brightness and leaves many structured residuals. The ePSF and \texttt{PSFex} models include structural details, resulting in unstructured, though still implausibly large residuals. Our Bayesian posterior sample gives detailed structure and pure noise-like residuals. Yellow pixels are masked for various reasons described in the text. A p-value test on the residuals (described in the text) gives 0.08 for our posterior sample, while for the other algorithms it confidently rejects the null hypothesis ($6\times 10^{-17}$, $2\times10^{-22}$, and $4\times10^{-13}$ respectively).}
    \label{fig:money}
\end{figure*}

In this work, we tackle the central challenge of representing a prior over PSF models explicitly. 
We replace the simple parametric, regularized, or implicit priors with a highly expressive, pixel-level prior $P(M)$ learned from high resolution sample images. 
To define our prior, we leverage score-based generative diffusion models~\citep{Song2021}, a framework seeing explosive growth in astrophysics~\citep[e.g.,][]{Adam2022, Mudur2022, Remy2023, Feng2023, Wu2024, Legin2025, Dia2025, Adam2025}. 
This approach enables, for the first time, robust sampling of the full, pixel-level posterior distribution $P(M | D)$, yielding substantially improved PSF reconstructions and a principled propagation of morphological PSF uncertainty to downstream analysis.
In \Fig{money} we demonstrate our main comparison, our PSF posterior samples give visually realistic models and superior, noise-like, residuals.
A final challenge, which we comment on but do not solve, is the generalization to a posterior $P(M(\vec{x})|D)$ of PSF models over the full detector frame rather than simply at the position of individual point sources.

In \Sec{data} we will present the Hubble-Space-Telescope (HST) data used for a concrete example of our analysis framework.
In \Sec{likelihood} we explain how the collected data forms a likelihood $P(D|M)$ which is shared over all methods considered here.
In \Sec{old} we examine three widely used traditional methods for modeling PSFs.
In \Sec{bayes} we present our fully Bayesian method for PSF reconstruction and demonstrate the robustness of the results.
In \Sec{discussion} we compare the various methods and comment on the applicability of our work.

\section{Data}
\label{sec:data}

While the methodology developed in this work is applicable to any astronomical imaging dataset, we demonstrate its efficacy using Hubble Space Telescope (HST) Advanced Camera for Surveys / Wide Field Channel (ACS/WFC) F814W filter images\footnote{This specific choice is motivated, though separate from, our other science goal: the high-fidelity analysis of galaxy-galaxy strong gravitational lenses from the Sloan Lens ACS (SLACS) Survey~\citep{slacs2006}. Our PSF models will be subsequently leveraged in \citet{Legin2025b}}. 
To ensure the most accurate noise modeling and robust likelihood calculation for our fully Bayesian approach, we operate exclusively on the flat-field calibrated, unstacked \emph{flt} frames. 
This is a deliberate choice not to use the common image stacking technique of Drizzle~\citep{Fruchter2002}.
Since every observation is distorted by a unique PSF (e.g., via focus breathing over an HST orbit), Drizzle would be combining structurally different PSF models to make a stack.
While it is common to use dithered images to improve sampling of the PSF to recover high resolution models, in such analyses care is taken to ensure the combined data share similar properties (detector position, orbital position, temperature, etc.).
This is not the case for our \emph{flt} frames and so stacking before modeling would be ill advised.
The dataset comprises 108 individual \emph{flt} frames, corresponding to multiple exposures across the selected fields\footnote{See \citet{Legin2025b} for a discussion of the selected targets from the SLACS survey corresponding to the \emph{flt} frames in question.}. 

Point source identification was performed using the matched filtering algorithm: IRAF StarFinder~\citep{Tody1986}, as implemented within the \texttt{photutils} framework~\citep{Bradley2025}. 
This routine identifies candidates by comparing a simple Gaussian kernel and then applying filters based on metrics of roundness and sharpness to isolate compact, unresolved sources. 
We used default selection values for the star finder, except for the Gaussian kernel we use a FWHM of 3 pixels, and for the detection threshold we used 50 times the background noise level\footnote{Determined as half the 16-84th percentile range of non-masked pixel values in the image.} in the image.
We experimented with a number of sharpness/roundness criteria, but ultimately decided to maximize completeness by moving all identified point source candidates to the visual inspection step.
Many small extended objects and artifacts were selected in the initial IRAF Starfinder run alongside true point sources.
We performed a visual inspection as a quality control step, selecting for isolated sources with no visible extended features.
After fitting (see \Sec{old} and \Sec{bayes}) a small number of extended object contaminants (75 sources) became apparent in the residuals\footnote{Visible in the residuals from multiple methods, not just our posterior samples.} and so these objects were discarded as well.
No other criteria were applied (i.e., signal-to-noise cuts), though our detection threshold for the IRAF Starfinder would have excluded exceptionally faint objects.
This process reduced the initial 3,573 identified candidates to a final working sample of \numps high-quality point sources analyzed in this work.

Each frame is accompanied by the requisite metadata, including a bad pixel mask identifying regions affected by detector artifacts, hot pixels, and cosmic rays. 
We implement a conservative masking strategy to ensure data quality. 
Specifically, any cosmic ray identified by the pipeline that spans more than one pixel is dilated by a single pixel to account for charge deposition in adjacent pixels, a phenomenon we observed in test analysis runs. 
All pixels exceeding 30,000 electrons are also masked, this is the empirical threshold for potential saturation in these data. 
The uncertainty associated with each pixel measurement is derived from the standard deviation provided in the \emph{ERR} map. 
This is converted to the corresponding variance map (the square of the \emph{ERR} map) for our Gaussian likelihood calculation under the assumption of independent pixel measurements.

\begin{figure*}
    \centering
    \includegraphics[width=\linewidth]{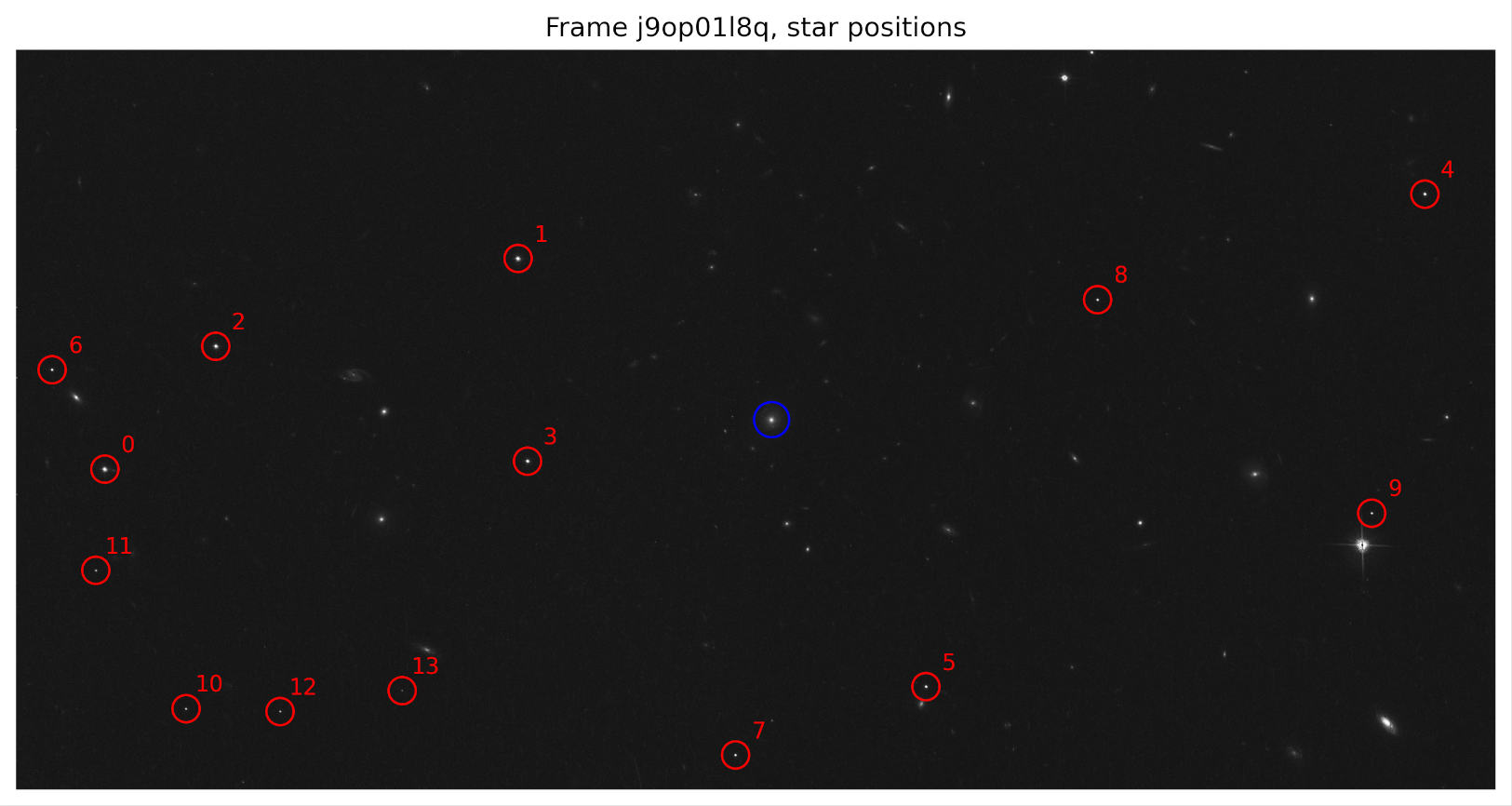}
    \includegraphics[width=\linewidth]{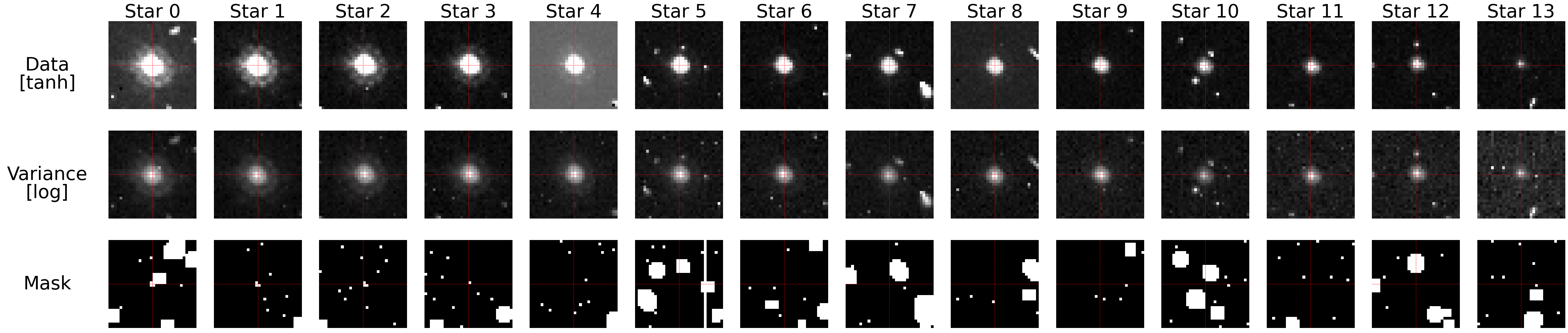}
    \caption{Example data from a single HST \emph{flt} frame. Top: The full frame, with selected stars (red) and the science target (blue). Bottom: Cutouts of the data, variance, and mask for each star. These cutouts illustrate the core challenge: each star presents a unique, complex PSF morphology that must be modeled from noisy, incomplete (masked) data.}
    \label{fig:data}
\end{figure*}

For each of the vetted point sources, we extract a $32\times 32$ pixel cutout.
This cutout size corresponds to a $128\times 128$ model image at four times upsampled resolution, which was set by the template ePSFs (see \Sec{epsf}) and computational considerations for our diffusion model prior (see \Sec{prior}). 
Each cutout is treated as a distinct realization of the PSF. 
To support the subsequent Bayesian modeling, a detailed set of associated metadata is stored, including the matching variance and mask cutouts, sky background level, the World and detector pixel coordinates of the cutout center, and the image exposure time. 
Before analysis, the raw flux in units of electrons is converted to normalized flux units (electrons per second) by dividing the data cutout by the exposure time. 
Similarly, the variance map is normalized by dividing by the square of the exposure time. 
\Fig{data} illustrates an example flt frame and a collection of extracted star cutouts\footnote{Note that star $\#10$ in the \Fig{data} cutouts was later dropped due to a difficult to see interloper, therefore later figures index the stars without star $\#10$.}.

\section{The Likelihood}\label{sec:likelihood}

We adopt a standard Gaussian likelihood based on the assumption of independent, normally distributed noise in each pixel.
The probability of observing the data vector $D$ given a set of model parameters is defined as:

\begin{equation}\label{equ:likelihood}   
    \mathcal{L}(D|M, x, y, f, s) = \mathcal{N}(D|f\cdot \mathbf{F}(M, x, y) + s, \Sigma), 
\end{equation}

where $D$ is the vector of flux values corresponding to the unmasked pixels in the data cutout. 
$M$ is a PSF model which may be evaluated at arbitrary pixel coordinates (it could be written as $M(i,j)$) even fractional pixel coordinates.
For most models under consideration, $M$ is a four-times-upsampled grid of pixels which are converted to a smooth model via bilinear interpolation; the parametric Moffat model is intrinsically defined at any position.
The set of nuisance parameters includes the relative sub-pixel shift vector $(x,y)$, the total source flux scalar $f$, and the uniform sky background $s$. 
$\Sigma$ is an $N_{unmasked}\times N_{unmasked}$ diagonal covariance matrix, where the diagonal elements are the variance values derived from the HST \emph{ERR} map for each unmasked pixel. 
By excluding masked pixels from $D$ and $\Sigma$, the mask is implicitly incorporated into the likelihood calculation.

Having a diagonal covariance matrix is further reason not to use the Drizzle algorithm, as the stacking procedure produces highly covariant noise in the stacked pixels.
While it is possible to model the covariance due to Drizzle, this would necessitate a dense $\Sigma$ matrix as opposed to merely using the diagonal, making the technique less computationally efficient.
Some pixel covariance may still be present in the \emph{flt} images (e.g., due to charge diffusion), though it was small enough not to noticeably affect our results in \Sec{bayes}.

The core of the forward model is the operator $\mathbf{F}$, which maps the high-resolution PSF model $M$ to the detector-scale observations. 
For each observed data pixel, $\mathbf{F}$ creates a $16\times 16$ grid of points and evaluates the shifted model $M$ at those locations.
Averaging those $16^2$ (256) samples gives the integral of flux in the data pixel contributed by the PSF model.
This high resolution integration is motivated by the need to account for sub-pixel centering effects crucial for high-precision astrometry and photometry.

This likelihood formulation serves as the fundamental objective function for all subsequent analyses, including both traditional and fully Bayesian approaches. 
The form of our likelihood is highly standard in the field of astronomical image analysis.
Further generalizations which include covariance between pixels do exist, such as the SLIC method~\citep{Legin2023}, though we found no evidence that such techniques were necessary in our data.

\section{Traditional Methods}\label{sec:old}

\subsection{Parametric PSF Modeling}\label{sec:parametric}

\begin{figure}
    \centering
    \includegraphics[width=0.9\columnwidth]{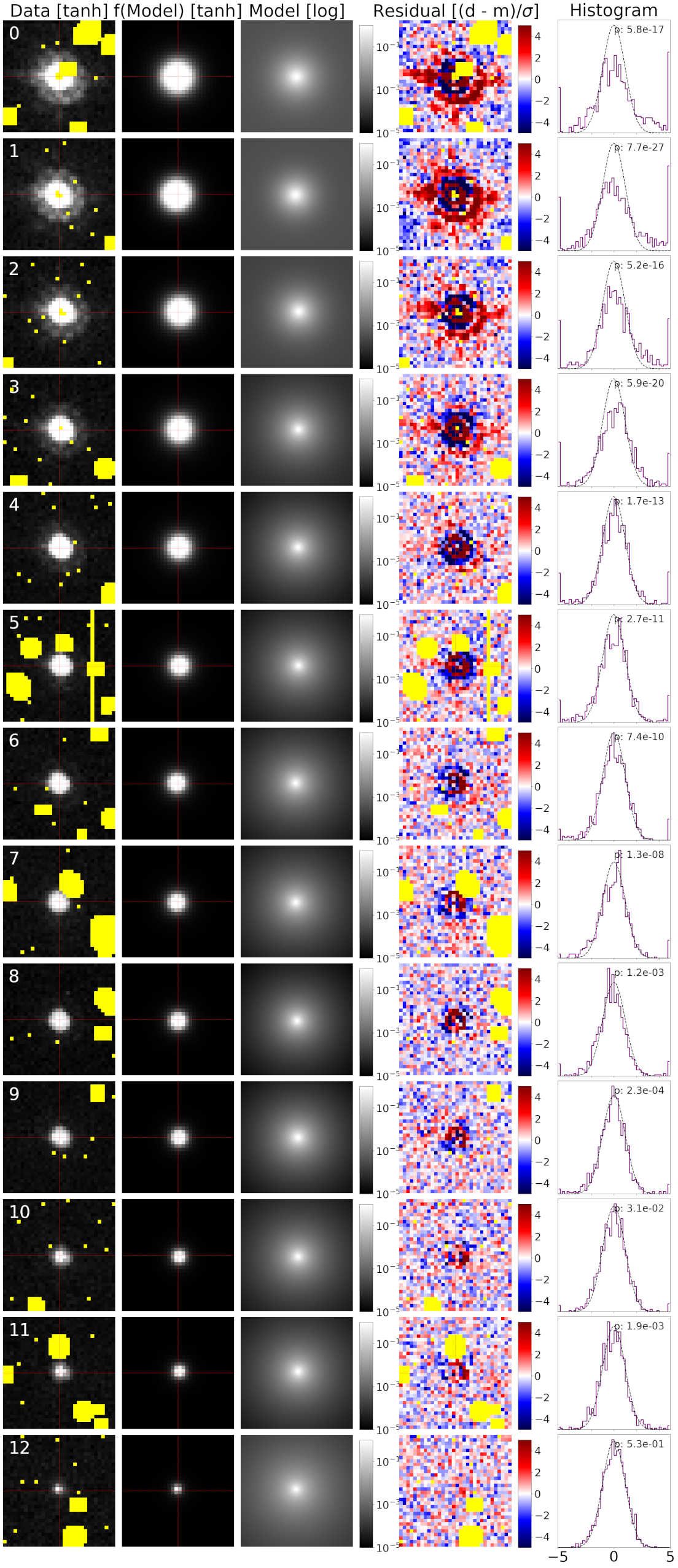}
    \caption{Traditional Method 1: Parametric Moffat Fit. From left to right the columns show the data, the best-fit Moffat model, the $4\times$ high-resolution model, the residuals, and a histogram of the residuals. Yellow pixels in the data and residuals are masked pixels (see text). The p-values in the top right of each histogram are from a Kolmogorov-Smirnov test on the central residuals against a Normal distribution~\citep{Virtanen2020}. The key takeaway is that this method fails: the strong, structured patterns in the residuals (column 4) and the non-Gaussian residual histograms (column 5, with p-values $\approx 0$) prove that a simple analytic function is a misspecified model for the true PSF.}
    \label{fig:astrophot}
\end{figure}

To establish a baseline for comparison, we first explore a standard parametric approach to PSF modeling. 
We utilize the Moffat profile~\citep{Moffat1969}, a common analytic form that is highly effective at characterizing the core structure of a PSF, although it lacks the complexity to capture high-order features such as diffraction spikes. 
The fitting is performed independently on each cutout using the general-purpose astronomical modeling code, \texttt{AstroPhot}~\citep{Stone2023}. 
In this implementation, the model consists of a single Moffat profile defined by its five parameters (centroid x and y, scale length, shape index, and amplitude), alongside a constant sky background as an additional nuisance parameter. 
The optimization is executed by minimizing the negative log-likelihood, which is computed for all unmasked pixels using the Gaussian formulation defined in \Sec{likelihood}.

The resulting fits are visually summarized in \Fig{astrophot}. 
While the simple Moffat profile successfully models the general structure of the PSF, it leaves pronounced systematic residuals. 
For the brightest sources, these residuals frequently exceed $5\sigma$ significance, which effectively rules out the possibility that the remaining structure is attributable to Gaussian noise.
This conclusion is quantitatively supported by the residual histograms in \Fig{astrophot}. 
A Kolmogorov-Smirnov test is performed on the central region of the residuals and yields vanishingly small p-values.
The central region residuals correspond to the most pronounced deviations; they are selected by dividing the cutout into a $3\times 3$ grid and taking the central segment.

Despite the significant systematic residuals, the formal uncertainties computed by \texttt{AstroPhot} for the Moffat parameters are exceedingly small. 
This result is not indicative of high accuracy but is a direct consequence of severe model misspecification. 
With only six parameters being constrained by approximately one thousand unmasked pixels, the model is highly overdetermined. 
Consequently, these small formal errors are misleading and do not provide a reliable measure of the true uncertainty associated with the real-world PSF.

\subsection{ACS/WFC Focus-Diverse ePSF Models}\label{sec:epsf}

Given the inherent limitations of the six-parameter Moffat model, a more flexible, pixelized representation is required to capture the detailed structure of the PSF. 
The effective PSF (ePSF) library developed by \citet{Anderson2006} provides such a model collection. 
An ``ePSF'' model combines the ideal instrumental distortion with detector-level effects, such as the pixel response and charge diffusion, yielding the final projected PSF necessary for most science cases.
The library was constructed by stacking thousands of high signal-to-noise stars and enforcing smoothness criteria across the spatial grid. 
Each template is a $101\times 101$ pixel model defined at a $4\times$ supersampled resolution (10,201 free parameters per template). 
To account for spatial variation, the library provides a grid of $9\times 10$ templates across the detector.

The PSF is subject to time-dependent variations, including focus changes driven by the Hubble Space Telescope's orbital thermal cycles \citep{Bellini2018}. 
To account for these perturbations, subsequent work has provided methods to derive an image-specific, customized template set, facilitated by easy-to-use web tools~\citep{Anand2023}. 
We adopt the \citet{Anand2023} template set specific to each frame for our analysis. 
Thus all \numps point sources are given a unique interpolated template to use for fitting.

We then fit these fixed templates to our stellar cutouts. 
This is performed using Maximum Likelihood Estimation (MLE), minimizing the negative log-likelihood defined in \Sec{likelihood}. 
For each cutout, only the four nuisance parameters are allowed to vary: the sub-pixel shift (x,y), the total flux (f), and the sky background (s). 
The optimization converges rapidly given the limited number of free parameters.

\begin{figure}
    \centering
    \includegraphics[width=0.9\columnwidth]{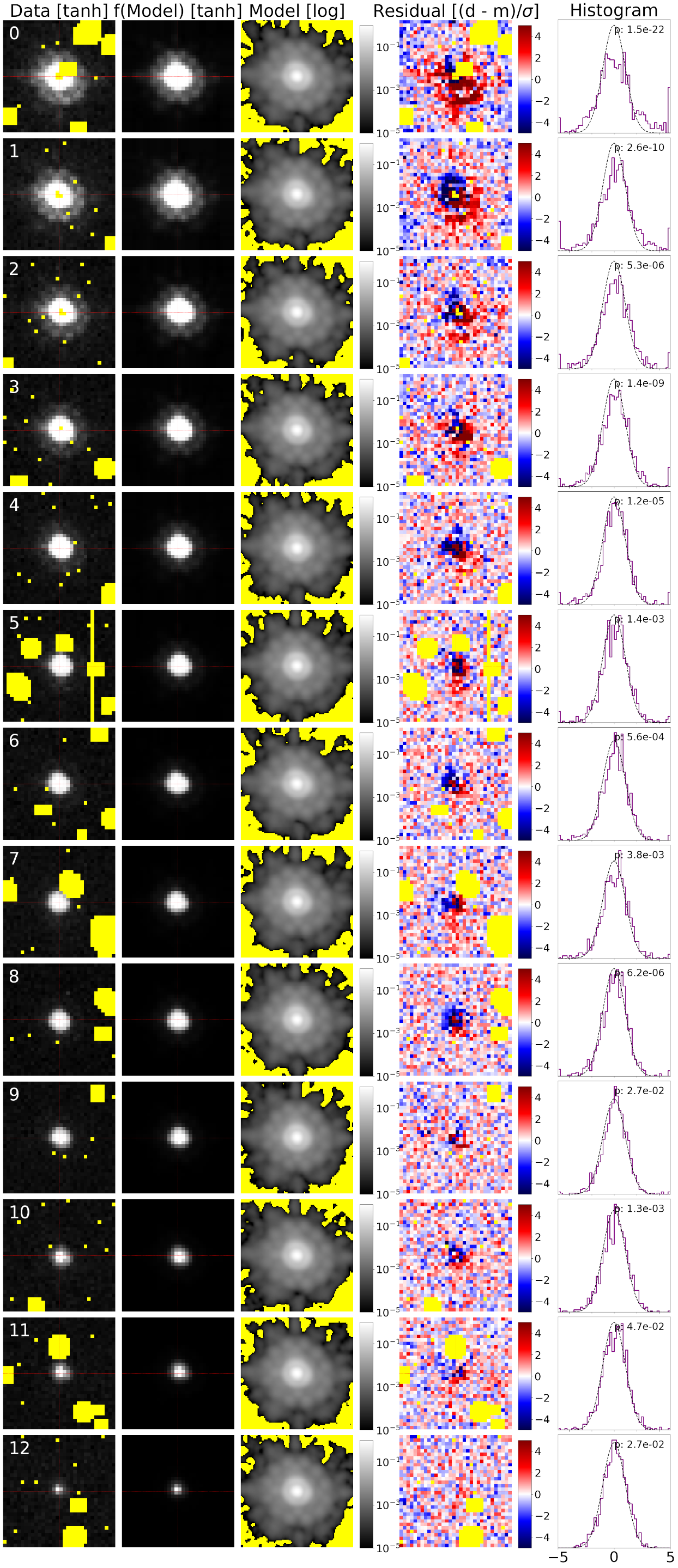}
    \caption{
    Traditional Method 2: Fixed ePSF Template. The format is identical to \Fig{astrophot}, yellow pixels in column 3 indicate pixel values below $10^{-5}$. This represents an improvement: the ePSF model (column 2) captures complex features like diffraction spikes, resulting in smaller residuals than the Moffat model. However, significant $>5\sigma$ residuals remain, proving that a single, fixed ``point estimate'' template is still an imperfect match to the unique PSF of any given star.}
    \label{fig:epsf}
\end{figure}

As demonstrated in \Fig{epsf}, the ePSF templates represent the observed cutouts with significantly greater fidelity than the simple Moffat profile. 
The most prominent systematic residuals, such as diffraction spikes and ring structures, are successfully removed by the more detailed model. 
However, the residuals still contain numerous pixels exceeding $5\sigma$ significance, especially in the PSF cores. 
This indicates that the fixed templates, while broadly effective at modeling HST PSF structures, do not perfectly represent the true PSF for a specific observation.

Fundamentally, this approach provides only a point estimate of the PSF structure. 
Since the template itself is fixed, the fit yields only formal uncertainties on the four nuisance parameters, failing to propagate uncertainties in the PSF's morphology.
This leads to biased and overconfident results even for the nuisance parameters which would be used in downstream analysis (astrometry and photometry).
This lack of model uncertainty coverage motivates our full Bayesian methodology.

\subsection{Regularized Pixelated PSF Modeling}\label{sec:psfex}

While template-based ePSFs offer improved accuracy, their reliance on external imaging campaigns means they are not perfectly tuned to the specific instrumental state of a given frame. 
We now investigate a method that performs pixelized optimization directly using the on-frame stellar observations. 
We adopt \texttt{PSFex}~\citep{Bertin2011}, a widely used algorithm designed to produce an empirical, spatially-varying PSF model by performing regularized pixel-level fitting to multiple point source cutouts simultaneously.

The \texttt{PSFex} workflow begins with source detection via \texttt{SourceExtractor}~\citep{Bertin1996} to identify candidate stars. 
We use the default \texttt{SourceExtractor} configuration except to set an appropriate initial PSF FWHM, pixel size, and desired cutout size.
\texttt{PSFex} then performs a $\chi^2$ minimization to determine the pixel values in a supersampled model, constraining the model's spatial variation across the detector with a low-order polynomial.
Critically, because this is a severely ill-posed problem (fitting thousands of pixel values from a limited number of stars), \texttt{PSFex} employs Tikhonov regularization~\citep{tikhonov1963} to stabilize the solution. 
This regularization acts as an implicit, simple prior, penalizing pixel values that deviate from a smooth model. 
We use the default \texttt{PSFex} configuration, except to set a $4\times$ supersampled model resolution and a first-order polynomial for spatial variation\footnote{higher-order models were observed to overfit our data given the limited number of stars ($\sim13$ on average) in each field}.

\begin{figure}
    \centering
    \includegraphics[width=0.9\columnwidth]{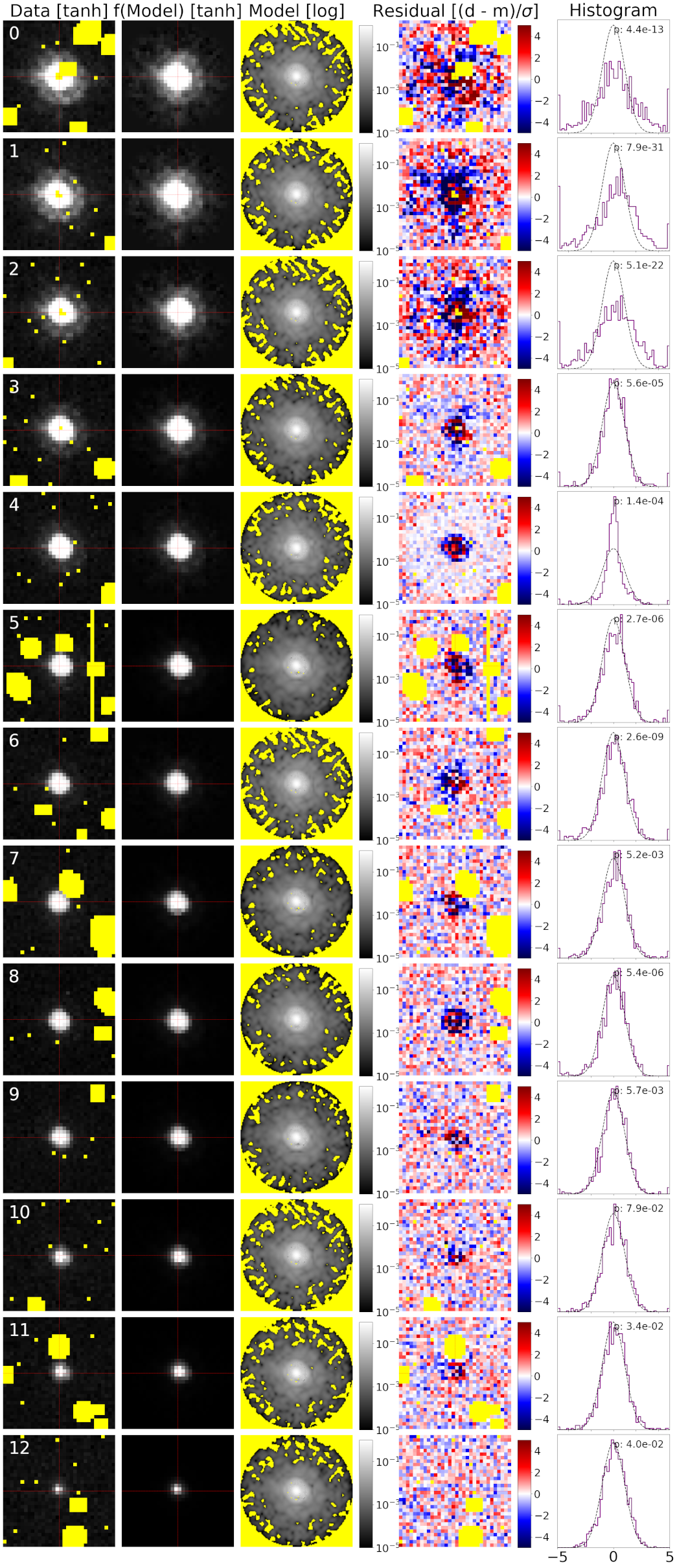}
    \caption{Traditional Method 3: Empirical \texttt{PSFex} Model. Format is identical to \Fig{astrophot}. This method shows mixed failure: most stars have implausibly large residuals, while others (e.g., Star $\#4$) have residuals that are too small, indicating the flexible model has overfit the data. This demonstrates that \texttt{PSFex}'s simple smoothness regularization is a misspecified prior, motivating our new Bayesian approach.}
    \label{fig:psfex}
\end{figure}

For direct comparison, we use the final \texttt{PSFex} model as a fixed template and perform a Maximum Likelihood Estimation for the four nuisance parameters (position, flux, and sky), identical to the procedure in \Sec{epsf}. 
The results are shown in \Fig{psfex}. 
While the overall fidelity appears comparable to the ePSF templates, the residuals for bright stars are often larger.
Disconcertingly, in some cases (e.g., Star $\#4$), the residuals are too small to be plausibly drawn from a Normal noise distribution, indicating that the flexible \texttt{PSFex} model has overfit the data. 
This overfitting is a direct consequence of the misspecified prior: Tikhonov regularization, which merely encourages smoothness, is a poor approximation for the true detailed structure of the HST PSF. 
Consequently, the fit does not yield the expected $\chi^2/\rm{DoF}\approx 1$ and, like the other traditional methods, provides only a single point estimate without a rigorous quantification of morphological uncertainty\footnote{Various techniques, such as bootstrapping, allow one to determine a distribution of solutions from methods like \texttt{PSFex}.
Though the formal meaning of these variational approaches is not always clear, they can give a good impression of the morphological variations in a PSF model.}.

\section{Pixel Level Bayesian PSF Modeling}\label{sec:bayes}

\subsection{Constructing a Prior PSF Model}\label{sec:prior}

\begin{figure}
    \centering
    \includegraphics[width=\columnwidth]{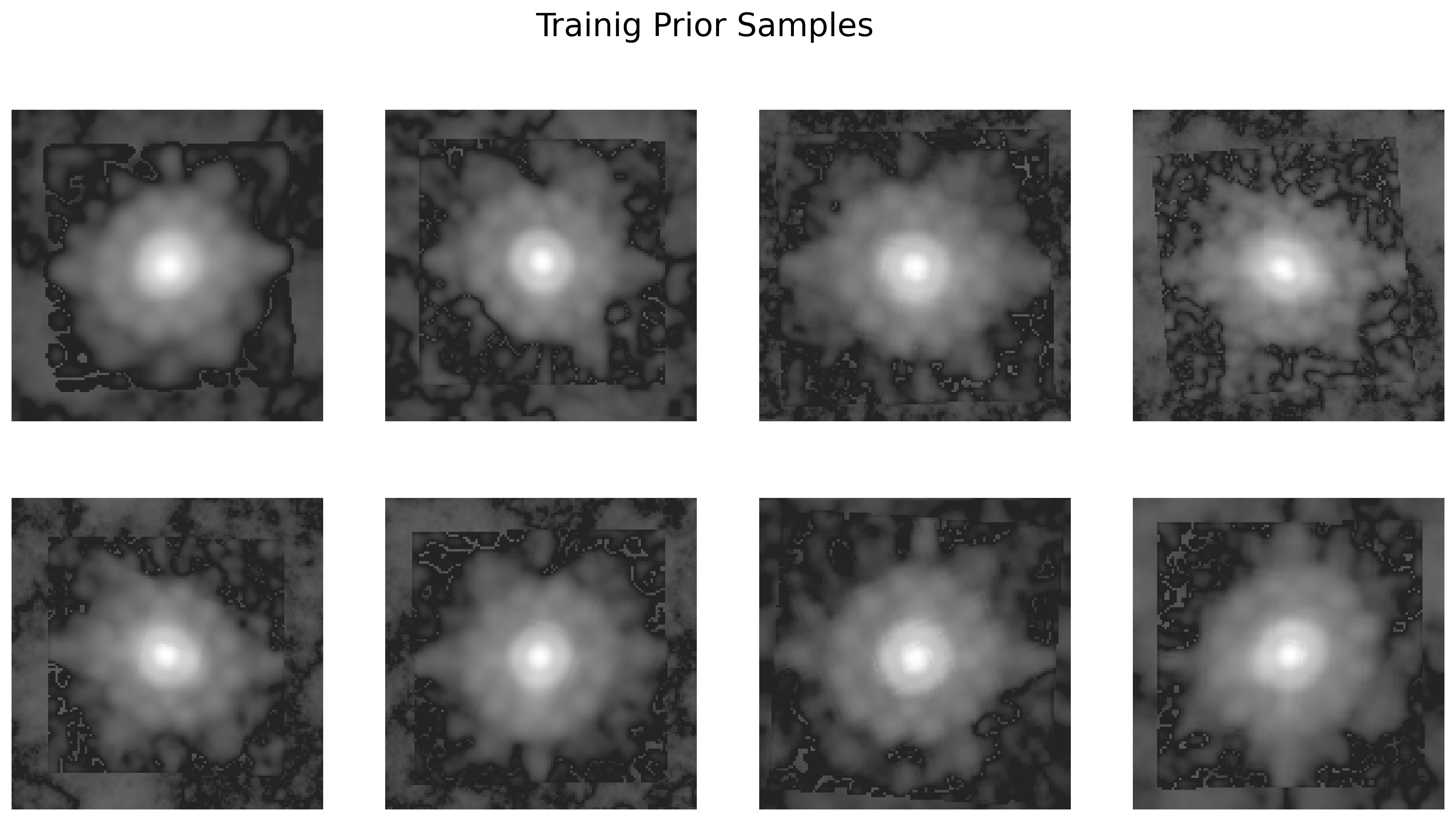}
    \includegraphics[width=\columnwidth]{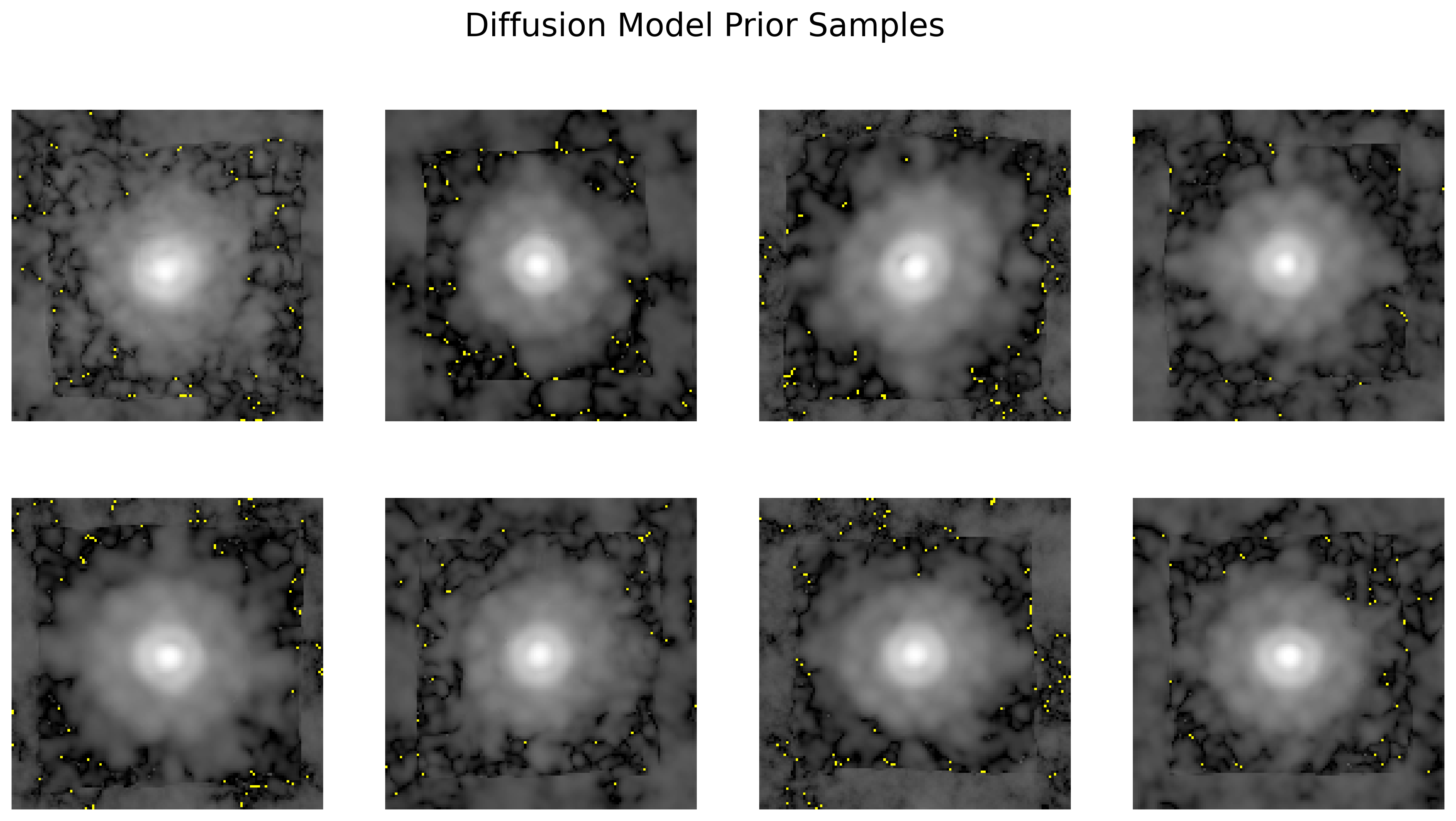}
    \caption{Top: Example prior samples based on the ePSF templates from the \citet{Anderson2006} library, used for training. Bottom: New, unique samples drawn from our trained diffusion model prior. The prior samples (bottom) are highly realistic, capturing the complex morphology of the training data (top), but are not simple copies. This demonstrates our model has learned a flexible, generative prior (a manifold of realistic PSFs) rather than just memorizing the training set.}
    \label{fig:prior}
\end{figure}

To perform a Bayesian analysis, we must first define a robust prior distribution, $P(M)$, over the space of plausible high-resolution PSF models $M$. 
All traditional methods, as discussed in \Sec{old}, can be understood as imposing a simple, and ultimately misspecified, prior.
Here, we instead construct a highly expressive, non-linear prior by training a generative model to learn the complex morphological features of realistic PSFs. 
We leverage score-based generative diffusion models~\citep{Song2021}, which implicitly represent this complex prior through a learned ``score function'', $\nabla_M \log(P(M))$. 
The application of diffusion models to problems in astronomy has seen recent explosive growth~\citep[e.g.,][]{Huang2018, Karchev2022, Adam2022, Mudur2022, Remy2023, Feng2023, Bourdin2024, Legin2024, Wu2024, Legin2025, Angeloudi2025, Dia2025, Adam2025}. 
This framework allows us to sample the posterior $P(M|D)$ by combining the log likelihood gradient, $\nabla_M\log(P(D|M))$, with this learned score function of the prior.

A key requirement for this approach is a representative training set. 
We use the 9,776 individual ePSF templates described in \Sec{epsf}. 
Given we showed in \Sec{epsf} that the ePSF templates match individual cutouts imperfectly, training on them may seem an odd choice. 
However, our premise is that while a single template fails, the full ensemble of templates robustly spans the manifold of realistic PSF morphologies (including focus, spatial, and thermal variations).
A further benefit of using score-based diffusion models is that they produce conservative priors with support outside their training samples; this allows for out of distribution posterior samples when the likelihood is highly informative~\citep{Adam2022, Rozet2024, Wagner2025, Barco2025}.
Our goal is to learn a prior that is broader than the training set. 
We achieve this in two ways. 
First, we apply extensive data augmentation to the training samples, including shifts, flips, scaling, rotations, skews, and blurring, to expand the effective prior space (each applying to $50\%$ of the training samples). 
Second, as we will discuss, we intentionally stop the model's training early to prevent it from ``memorizing'' the discrete templates.

Our diffusion model, implemented using a U-Net architecture~\citep{Ronneberger2015} from the \texttt{score\_models} package\footnote{\href{https://github.com/AlexandreAdam/score_models}{https://github.com/AlexandreAdam/score\_models}}, has several key technical features. 
For numerical efficiency, the $101\times 101$ templates are padded to a $128\times 128$ resolution; the empty outskirts are filled with a Gaussian Random Field whose power-law spectrum approximately matches the observed PSF outskirts. 
To enforce the physical constraint of non-negativity, all training models are thresholded at a floor value of $5\times 10^{-5}$ relative to the peak flux. 
This value was chosen to be safely above the minimum noise level ($1\times 10^{-5}$) used in the diffusion model's sampling process, again ensuring numerical stability.

Given the high dynamic range of the PSF (over $10^5$), we found it numerically unstable to model the full pixel values directly. 
Instead, we compute an average template $\bar{M}$ and train the diffusion model to learn only the deviations from this mean. 
This dynamic-range-reducing transformation is applied purely for numerical stability and has no effect on the validity of our analysis (i.e., it applies without loss-of-generality).

Training duration proved to be an important hyperparameter. 
Extended training caused the model to overfit, concentrating its score function on the training examples. 
This ``spiky'' prior would then resist solutions favored by the likelihood, giving visible residual structures. 
By empirically selecting an ``early stopping'' point, we obtained a model that balances prior expressiveness with flexibility. 
The early stopping point was determined by training models with different numbers of epochs and comparing their resulting posterior samples. 
Models trained for a large number of epochs consistently produced visible residuals with their posterior samples. 
Models trained for too few epochs produced prior samples that typically appeared drawn from our input prior distribution, but included occasional ``catastrophic failures'' that were highly out of distribution. 
We adopted an intermediate number of training epochs that didn't suffer from the two identified failure modes.
Our results were robust to the exact choice of the number of training epochs and so we did not further optimize this hyperparameter.
As shown in \Fig{prior}, all the morphological features of the training data samples can be seen in our prior samples, meaning the overall structure of the PSF has been learned.
That said, samples drawn from our prior have slight visual distinctions from the training examples, confirming that we have defined a broader prior manifold.

A minor, yet notable, boundary artifact is visible in the extended wings of the prior samples, producing a faint square structure. 
This feature is directly inherited from the padding process used to expand the $101\times101$ templates to the $128\times 128$ input size for the U-Net. 
Critically, this structure occurs at fluxes approximately $10^{-4}$ relative to the PSF peak, it minimally affects our posterior sampling procedure. 
We recommend that the posterior samples be cropped back to $101\times 101$ pixels for further analysis after sampling.
Our analysis in \App{flux} found that the faint border could bias flux estimates by $\sim 1-2\%$ so we perform this cropping for all quantitative analyses\footnote{ In plots we leave the border in place for the sake of interest to the reader.}.
Alternatively, sampling on extremely high signal-to-noise sources allows the dominant likelihood term to effectively overpower the broad prior in the low-flux regimes, naturally correcting the artifact, as demonstrated in \citet{Adam2022}.

Several promising avenues exist for improving our diffusion model prior. 
Our current reliance on the \citet{Anderson2006} ePSF templates could be improved by training the model on high-precision wavefront PSF models~\citep{Perrin2012, Desdoigts2023}, which might provide a more physically motivated basis and extend to fainter features. 
Our prior model is also agnostic to the position, flux, or color/spectrum of the source being modeled.
Since the PSF is known to be a function of these properties, we could further narrow the prior model distribution by incorporating them into the model, rather than producing a single broader model that covers all cases.
These directions for improvement would be interesting directions for further progress, but are beyond the scope of this work.

\subsection{Posterior Samples}\label{sec:posterior}

\begin{figure}
    \centering
    \includegraphics[width=0.9\columnwidth]{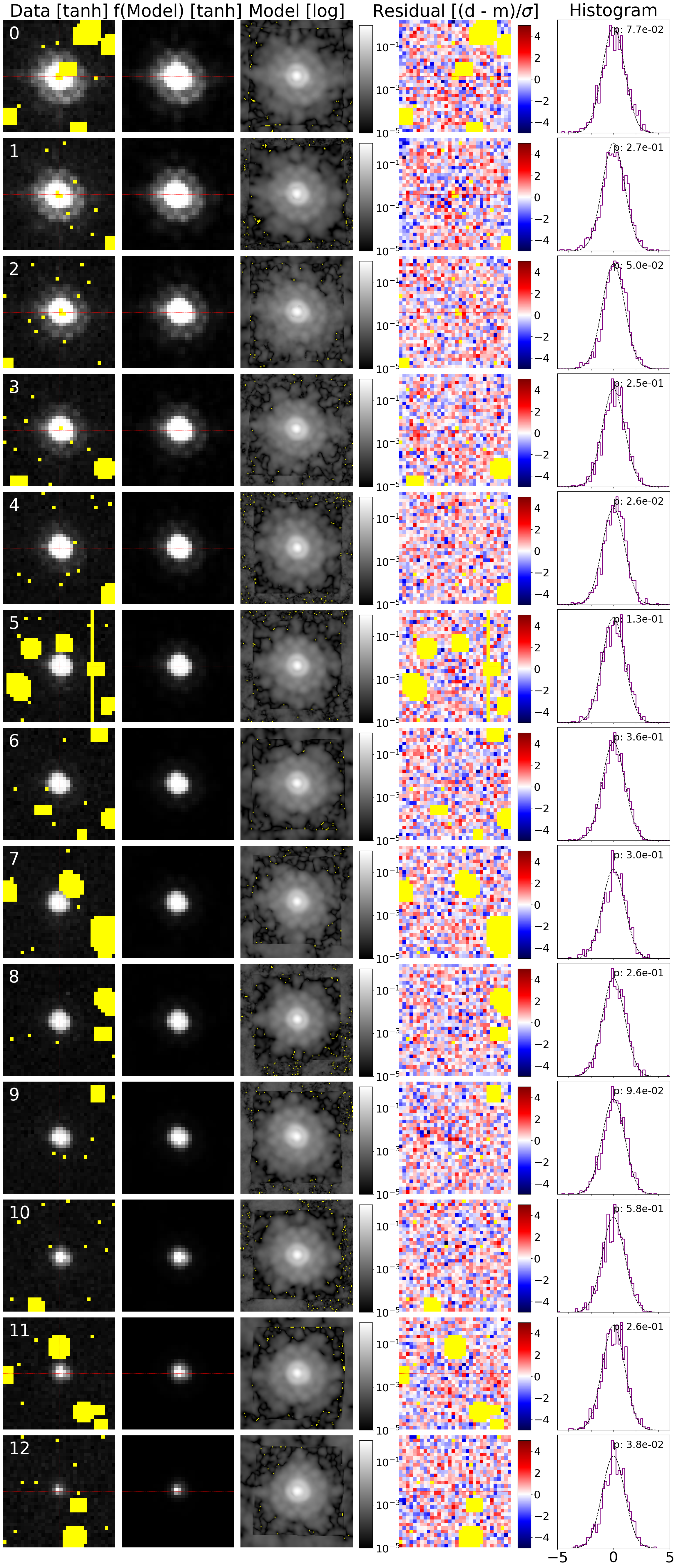}
    \caption{
    Pixel-level Bayesian posterior PSFs (our work). Format is identical to \Fig{astrophot}. The model shown (column 3) is a single, random sample from the posterior rather than a point estimate of the PSF. The residuals (column 4) are unstructured and consistent with pure noise, and the residual histograms (column 5) are consistent with a Normal distribution (KS test p-values often $\approx 0.5$). This demonstrates our framework produces a statistically sufficient model for the data.}
    \label{fig:posterior}
\end{figure}

Our diffusion model prior requires a linear forward model to enable efficient sampling~\citep{Song2021, Adam2022}. 
However, the full likelihood in \Equ{likelihood} is non-linear with respect to the combined sub-pixel shift $(x,y)$, flux $f$, and sky level $s$. 
We therefore cannot sample all parameters simultaneously.
Instead, we adopt a pragmatic, iterative approach that alternates between optimization and sampling:
\begin{enumerate}
    \item Optimize Nuisance Parameters: We hold the PSF model $M$ fixed and find the Maximum Likelihood values for the nuisance parameters $(x, y, f, s)$.
    \item Sample the PSF: We fix the nuisance parameters at their optimized values, and sample from the PSF posterior. 
\end{enumerate}
For the initial round we set the PSF model to $\bar{M}$ the average ePSF template sample.
The forward model $\mathbf{F}$ is now a simple linear operation mapping $M$ to the data in step 2. 
This linearity permits the use of the ``convolved likelihood approximation'' from \citet{Adam2022}, a computationally efficient method for diffusion model posterior sampling.
We cycle this two-step procedure twice to ensure all parameters have settled. 
We find this is sufficient, as any minor, remaining uncertainties in the nuisance parameters are effectively marginalized by the high-dimensional, pixel-level flexibility of the PSF posterior itself.
This is because the PSF posterior sampling has some small flexibility on the position, flux, and sky level by controlling all pixels.
Our PSF prior is trained with variability in flux and position; consequently, our PSF posterior has flexibility to explore variations in those parameters. 
The sky level has the least flexibility to explore outside of the nuisance parameters optimization step, but even that has some variability at the level of $10^{-5}$ (relative to the peak brightness) due to the sampling procedure.
From our nuisance parameter fitting, the median sky level deviation from the initial sky estimate is $8\times10^{-6}$, however some fits deviate by as much as $0.02$ in flux.
In a final run, we produce 32 posterior samples to represent the posterior distribution for each cutout.

This sampling process involves numerically integrating a Stochastic Differential Equation (SDE) defined by the combined prior score function and the likelihood gradient. 
Conceptually, this procedure begins with pure noise and iteratively ``de-noises'' it; at each step, the sample is guided by both the likelihood (what the data requires) and the prior (what a realistic PSF looks like), ensuring the final sample is a valid posterior sample.
For the integration, we employ the Heun SDE integrator from the \texttt{score\_models} package. 
We use 4,096 integration steps and incorporate two Langevin corrector steps per iteration\footnote{During the initial two cycles, where we optimize the nuisance parameters, 2,048 SDE steps and no corrector steps are used.}, which helps reduce discretization errors from the SDE integration.
An implementation and example of our posterior sampling procedure is available here: \href{https://github.com/Ciela-Institute/HSTPSF}{https://github.com/Ciela-Institute/HSTPSF}.

\Fig{posterior} presents a single, randomly chosen posterior sample for each cutout. 
The results are a clear improvement over the methods in \Sec{old}. 
The residual images exhibit essentially no discernible structure, and the $\chi^2/\rm{dof}$ is approximately one for all examples. 
This strongly indicates that our Bayesian framework has produced a statistically sufficient model, successfully absorbing all non-noise signal from the data into the posterior ensemble. 
While a faint residual is visible in cutout 1, this is not present in all posterior samples for that star, highlighting the importance of using the full posterior ensemble rather than a single realization.

\begin{figure}
    \centering
    \includegraphics[width=\linewidth]{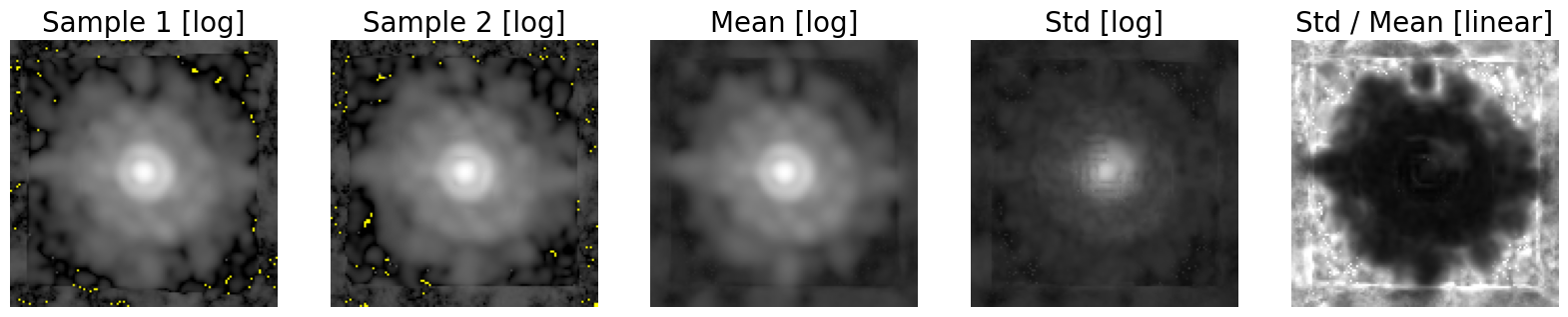}
    \includegraphics[width=\linewidth]{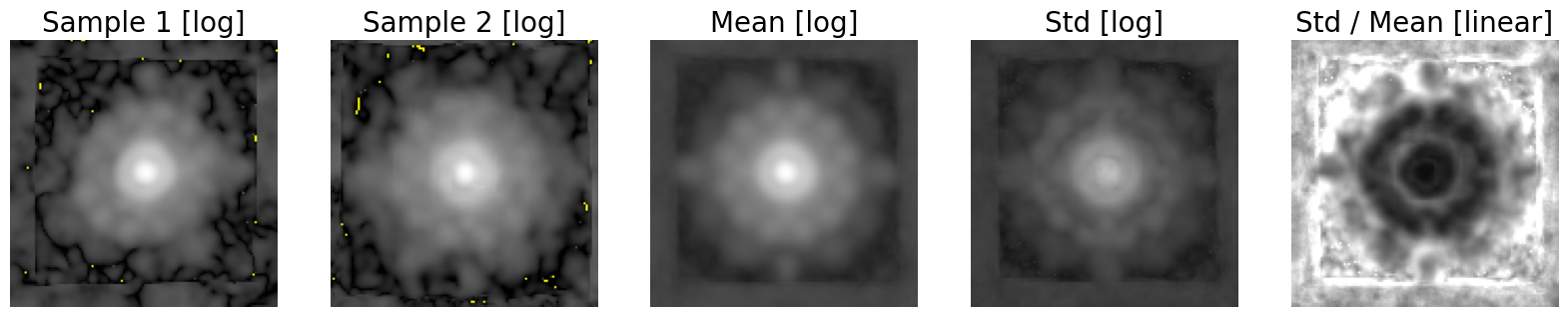}
    \caption{Variance of Posterior samples. Columns show two random posterior samples (1-2), the posterior mean (3), the standard deviation (4), and fractional standard deviation (5). These are based on 32 posterior samples per cutout. This figure demonstrates the power of our method: we now have a principled, pixel-level quantification of morphological uncertainty (which may be marginalized as in column 4). As expected, the bright star (top) is highly data-constrained (low variance), while the faint star (bottom) is less constrained (higher variance), producing a broader posterior.} 
    \label{fig:variance}
\end{figure}

\Fig{variance} visualizes the quantified uncertainty by showing the marginal variance map across the 32 posterior samples for two example sources. 
As expected, the high-SNR star (top) is more tightly constrained than the low-SNR star (bottom). 
For both, the bright central regions are data-constrained, while the faintest outskirts are almost entirely prior-driven. 
The posterior mean is shown for visualization, but we stress that the posterior mean is not a posterior sample and holds no special statistical meaning, as the true posterior is highly non-Gaussian.

\subsection{Interpolating to an image position}\label{sec:interpolate}

Our Bayesian procedure yields posterior samples $\{M_k\}$ at the discrete locations of observed point sources. 
However, most downstream analyses require a PSF model at an arbitrary position $\vec{x}$ within the detector plane (e.g., at the centroid of an extended galaxy). 
This necessitates a method for interpolating the PSF field.
Interpolation is a non-trivial challenge in PSF modeling. 
Standard methods, such as fitting a low-order polynomial across the detector (e.g., \texttt{PSFex}), are severely under-constrained given our typical source density of 2 to 20 stars per frame.
A more rigorous, high-dimensional Gaussian Process (GP) is computationally prohibitive~\citep{Rasmussen2003gaussian}, as our PSF model exists in a $D=16,384$ dimensional space\footnote{This is simply the number of pixels in the model: $128^2$.}. 
Furthermore, it is unlikely that our relatively small collection of 32 posterior samples per point source are sufficient to robustly estimate the high-dimensional covariance required for a meaningful GP model over the field.
While a fully Bayesian approach would involve a prior conditioned on detector position allowing for PSF posteriors over the full detector field, this is beyond the scope of this work.

Given the intractability of a rigorous interpolation, we adopt a pragmatic, non-parametric weighted mean approach. 
An interpolated PSF model $M(\vec{x})$ is computed as a weighted average of posterior sample draws from all observed sources, $\{M_k\}$. 
The weight $w_k$ for each source $k$ is defined as:
\begin{equation}
    w_k = \frac{1}{\sigma_k^2}e^{-0.5 \frac{||x_k-x||^2}{1024^2}},
\end{equation}
where $||\vec{x}_k-\vec{x}||$ is the spatial distance, $\sigma_k^2$ is the average variance of that source's posterior\footnote{This is an average measure of the result in the second from right column of \Fig{variance}. More well constrained posteriors will produce small $\sigma_k$ and poorly constrained posteriors will produce large $\sigma_k$.}, and the weights are normalized to sum to one. 
The spatial weighting kernel is intentionally broad; this ensured that the interpolation is approximately evenly weighted with the variance.

To propagate the posterior uncertainty, we employ a Monte Carlo approach: we randomly select one posterior sample from each source's ensemble, perform the weighted interpolation, and repeat this procedure multiple times. 
This generates a ``pseudo-posterior'' ensemble at the target position $\vec{x}$, capturing morphological variations.

While this weighted interpolation scheme is a practical compromise, we validate its performance using a leave-one-out cross-validation. 
In this test, we remove a star, interpolate a pseudo-posterior to its location using the remaining stars, and compare the result to the held-out data. 
The results of this validation, which demonstrate superior residuals to the methods in \Sec{old}, are presented in \App{interpolate}.

\section{Quantitative Comparison}\label{sec:comparison}

The above visuals are compelling, but would be complemented by a quantitative comparison.
Here we consider the reduced $\chi^2$ statistic, a commonly used goodness-of-fit measure.
For our number of degrees of freedom (DoF), we take the number of unmasked pixels.
An expectation for any ``good'' model is that the reduced $\chi^2$ will approximately equal one.
While often one subtracts the number of ``free parameters'' from the degrees of freedom, this concept is poorly defined for all but the Moffat model and is not appropriate in the Bayesian context we consider here\footnote{In the Bayesian framework, the true PSF model is sampled from a prior: $M_i \sim P(M)$, then the data is sampled from the likelihood: $D_i \sim P(D|M_i)$. Since our likelihood is Gaussian on the pixels, the expectation for any combination of normalized pixel residuals is mean zero, variance one, which is equivalent to $\chi^2$ of one.}.

\begin{figure}
    \centering
    \includegraphics[width=\linewidth]{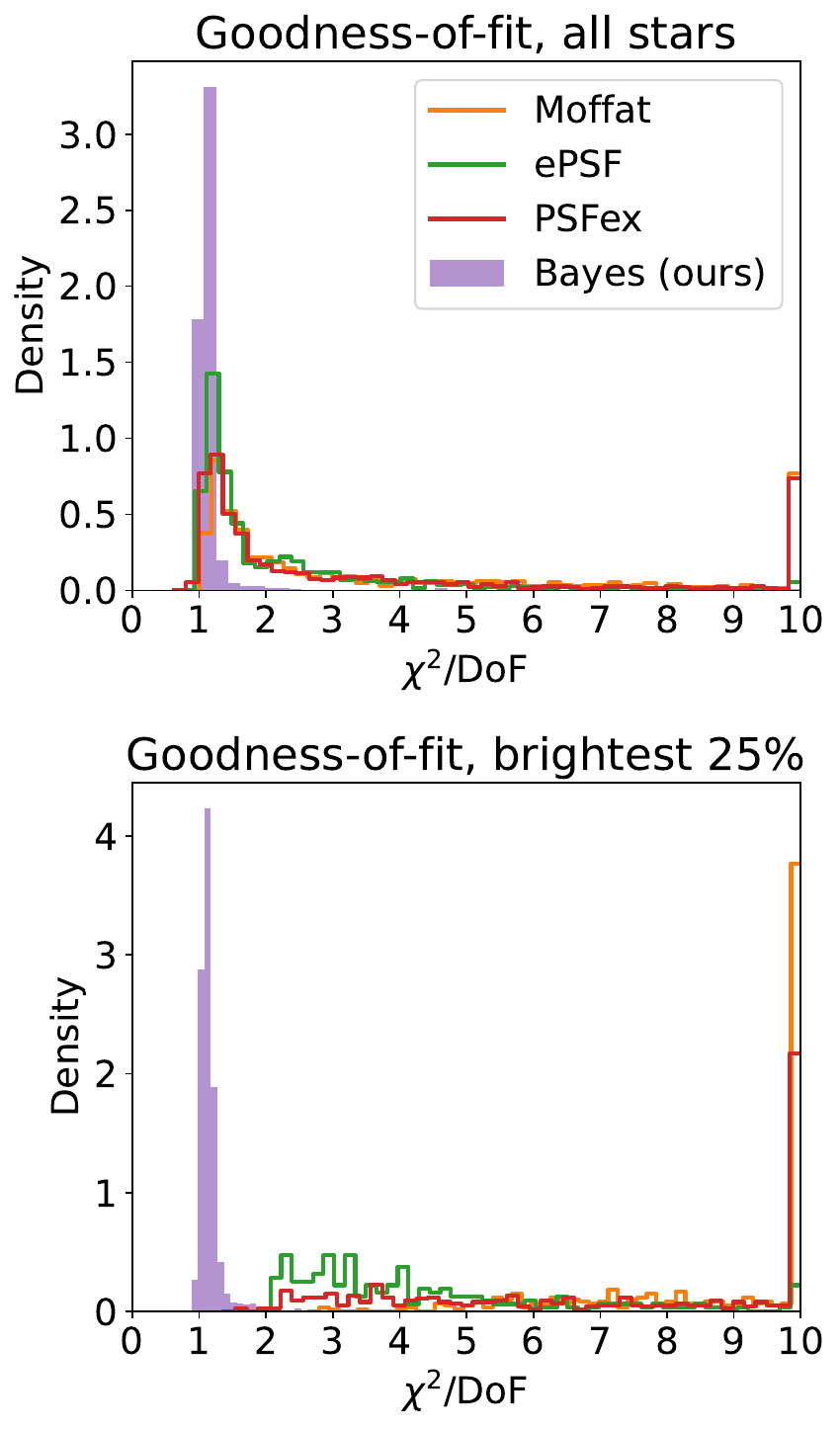}
    \caption{Goodness-of-fit for all \numps point sources for each method. Upper panel, gives the reduced $\chi^2$ step histogram (normalized to a density) for each method, where our Bayesian method is filled in (purple) for contrast. The histograms are truncated at $\chi^2/{\rm DoF}=10$ and any values exceeding this limit are placed in the last bin. Lower panel, same as upper panel except only for the brightest quartile of point sources. The Bayesian method produces 32 posterior samples per point source, we include all these samples in the histogram for completeness; the other methods produce only one model per point source.} 
    \label{fig:comparison}
\end{figure}

\Fig{comparison} presents the reduced $\chi^2$ for all \numps point sources as fit by each method.
The upper panel gives all stars, while the lower panel gives the brightest 25\%.
As noted throughout our discussion, the faint stars are relatively easy to fit, with even the highly oversimplified Moffat model performing reasonably well on the faintest stars.
By selecting the brightest quartile, we truly see each method's relative performance.
As would be expected from our earlier visual comparisons, our posterior sampling procedure performs nearly identically in both panels, while all other methods struggle\footnote{We note again that \texttt{PSFex} is not being testing under its ideal conditions. With many point sources to fit, \texttt{PSFex} may model effects such as the flux dependence of the PSF.} ($\chi^2/{\rm Dof} \gtrsim 2$).
Observe that our Bayesian method improves on other methods both by average performance, and in reducing the number of ``catastrophic failures'' implying the method is more robust.
As our model is defined in pixel space, it is always able to adjust to accommodate variations in the high SNR point sources. 
More rigid models must make compromises which ultimately leave several pixels at many ($>5$) standard deviations from what the data implies naturally.

It is worth noting that while our posterior sampling procedure performs exceptionally well, the $\chi^2/{\rm DoF}$ distribution is not peaked at one.
This would be expected if our prior distribution was the true distribution from which HST PSF models are drawn.
Our inferred prior from modified ePSF templates is only an approximation of this ``ground truth prior'' and so there is room for improvement in this aspect of our analysis.
Perhaps taking a prior from samples of a wavefront model of the instrument would produce even better posterior samples that this quantitative test could readily detect as superior.
this is an exciting avenue for future exploration.

\section{Discussion and conclusions}\label{sec:discussion}

We have successfully demonstrated a novel framework that produces pixel-level Bayesian posterior Point Spread Functions (PSF) in HST imagery using a Gaussian likelihood and generative diffusion model prior.
This approach was benchmarked against three traditional PSF modeling techniques: simple parametric fitting (Moffat), fixed template libraries (ePSF), and empirical regularized likelihood methods (\texttt{PSFex}). 
In comparison, our method not only yields visually more realistic PSFs but, critically, achieves likelihood values orders of magnitude greater.
Our PSF models give residuals statistically consistent with noise in most cases and adequately capture the detailed structure of the effective PSF. 
The production of a full posterior distribution for the PSF morphology enables the previously intractable marginalization over model uncertainty in subsequent analyses. 
This is not merely a formal improvement; we demonstrate in \App{flux} that traditional methods can introduce systematic flux biases of order 3\%.
Our procedure also yields unbiased measurements for significantly fainter stars than traditional methods, simply producing a broader posterior in accordance with the lower signal-to-noise.
An implementation and example of our posterior sampling procedure is available here: \href{https://github.com/Ciela-Institute/HSTPSF}{https://github.com/Ciela-Institute/HSTPSF}.

One limitation of the work is the requirement of PSF training samples for the generative model used as prior.
In our case, previous works by \citet{Anderson2006} extracted highly detailed upsampled PSF models which we could adopt for training.
Another viable approach is to use analytically modeled PSFs with wavefront propagation~\citep{Desdoigts2023}.
In the absence of either type of pre-determined PSF models, another approach would be to learn the upsampled PSF model directly from the data using an iterative maximum likelihood algorithm such as described in \citet{Rozet2024, Barco2025}.

The primary limitation identified in this work is the use of a non-parametric, ad-hoc weighted mean for interpolating the PSF posterior to arbitrary image locations. 
Our chosen weighting scheme (based on posterior variance and spatial proximity) is computationally efficient and practical. 
A leave-one-out cross-validation test in \App{interpolate} demonstrates this interpolation scheme is superior even to the fitting results of traditional models, though it lacks the rigor of a full Bayesian formulation.
Integrating a principled interpolation scheme, such as a high-dimensional Gaussian Process, Neural Network, or a flexible polynomial expansion (like that used in \texttt{PSFex}), is an exciting direction for future work.
The most formal approach would be a PSF prior conditioned on detector position; given a likelihood at the position of individual stars, this would produce posterior realizations of the full frame variable PSF model.

Despite the current limitations in interpolation, the availability of full, quantified PSF posterior distributions holds transformative implications across a number of astronomical disciplines. 
As discussed in the introduction, this work directly enhances the precision of measurements in fields including weak gravitational lensing, astrometry, high-contrast imaging, galaxy scaling relations, and photometry. 
By providing a statistically complete description of the PSF uncertainty, this method introduces a new paradigm of precision PSFs which push to the limits of information in a given point source cutout. 
We anticipate that this work will enable levels of astrophysical precision previously considered intractable.

\section*{Acknowledgments}

C.S. acknowledges the support of a NSERC Postdoctoral Fellowship and a CITA National Fellowship.
This research was made possible by a generous donation by Eric and Wendy Schmidt via the Schmidt Sciences Foundation. 
Y.H. and L.P. acknowledge support from the Canada Research Chairs Program, the National Sciences and Engineering Council of Canada through grants RGPIN-2020- 05073 and 05102. N.M. acknowledges funding from CIFAR, Genentech, Samsung, and IBM.
The work of A.A. and R.L. were partially funded by NSERC CGS D scholarships. 
R.L. acknowledges support from the Centre for Research in Astrophysics of Quebec and the hospitality of the Flatiron Institute. 
G.M.B. acknowledges support from the Fonds de recherche du Québec – Nature et technologies (FRQNT) under a Doctoral Research Scholarship (\href{https://doi.org/10.69777/368273}{https://doi.org/10.69777/368273}).

Software used: \texttt{astropy}~\citep{astropy2022}, \texttt{AstroPhot}~\citep{Stone2023}, \texttt{caskade}~\citep{Stone2025a}, \texttt{matplotlib}~\citep{Hunter2007} , \texttt{numpy}~\citep{harris2020array}, \texttt{PyTorch}~\citep{Paszke2019}, \texttt{PSFex}~\citep{Bertin2011}, \texttt{SourceExtractor}~\citep{Bertin1996}, \texttt{scipy}~\citep{Virtanen2020}, \texttt{score\_models}~(Adam et al., in-prep.), \texttt{PTED}~\citep{stone2026pted}

\bibliography{diffusionpsf}{}
\bibliographystyle{aasjournalv7}

\begin{appendix}

\section{Ground Truth Test}\label{app:inject}

Our tests thus far have focused on real HST data where the true PSF model is unknown to us, here we perform a test in which the ground truth is a known PSF model.
To perform our test we take the posterior sample model PSFs shown in \Fig{posterior} and now set them as the ground truth PSF to generate new mock data.
We use the same variance maps, masks, etc. to reproduce a dataset and meta-data collection that is identical in every way except that the data itself has been replaced using a known PSF model.
We then run the exact same posterior sampling procedure as described in \Sec{bayes}.
This allows us to test how well our sampling procedure covers the ground truth under controlled conditions.
For a true coverage test, we would need to use samples drawn from our PSF prior model, however using posterior samples from real data makes for a more challenging test of the model as the posterior samples are potentially low probability samples from the prior.

\begin{figure}
    \centering
    \includegraphics[width=0.5\linewidth]{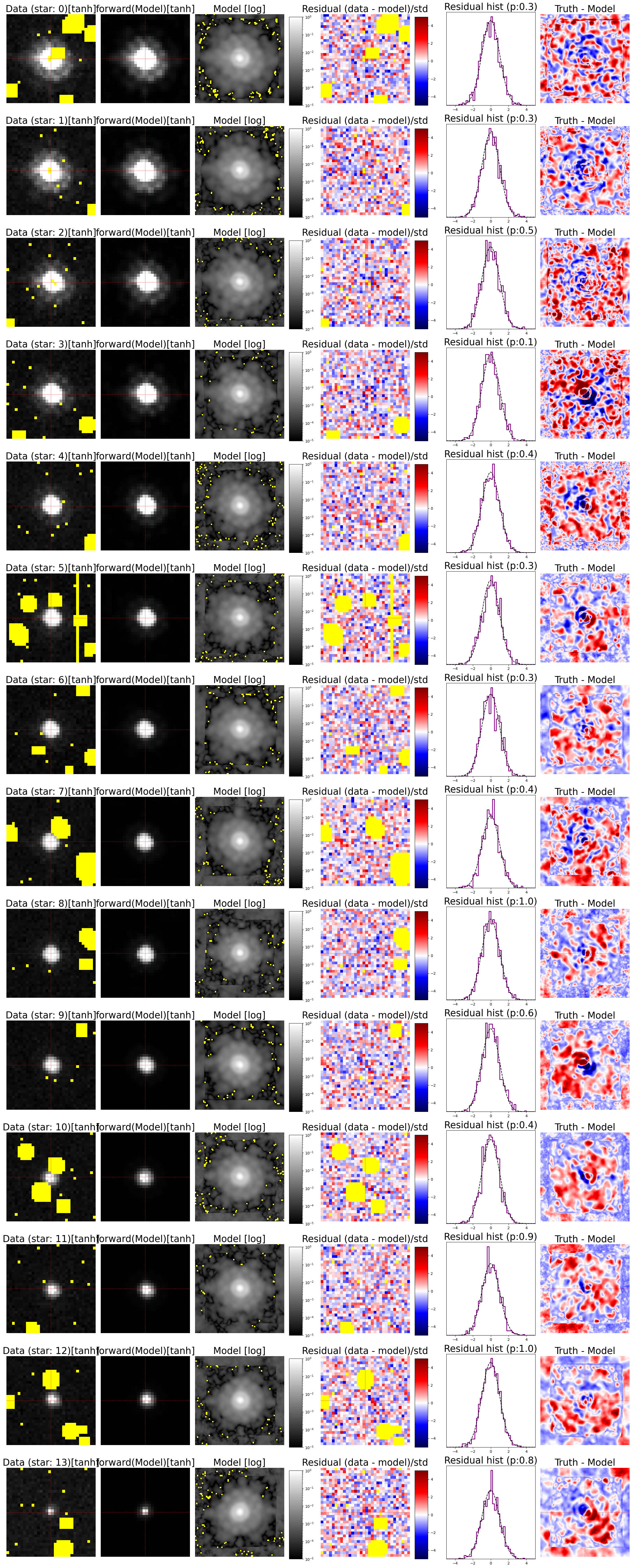}
    \caption{
    Injection test with known ground truth PSF model. Format is identical to \Fig{astrophot} except for the final column. The last column shows the difference between the ground truth PSF model and the mean posterior PSF model, normalized by the per-pixel standard deviation of the posterior PSF model. The color scale is the same as for column four.}
    \label{fig:inject}
\end{figure}

\Fig{inject} presents the resulting fits to our injected data test.
Here we see largely similar results to \Fig{posterior} for the first five columns, as should be expected given our ability to fit near identical data earlier.
The final column reveals interesting information on the capability of our posterior sampling procedure.
This column shows the residuals of the ground truth PSF model (at full $4\times$ upsampled resolution) compared with the mean of our posterior model samples, normalized by the posterior per-pixel standard deviation.
While this comparison has no explicit statistical meaning\footnote{For example, the mean of the posterior samples is possibly not contained in the posterior PDF. Similarly, the per-pixel variance of the posterior samples is not the true high-dimensional width of the posterior distribution. Therefore, direct pixel-wise comparison between the marginal posterior variance and the ground truth is purely illustrative.}, we may gain insight from these comparisons.
The first notable feature is that the ground truth vs posterior residuals are broadly within a few ``sigma'' of the posterior.
The color scale we use is the same as in column four and so only saturated red/blue pixels show highly over-sized deviation; of which there are very few.
The most significant deviations occur for star $\#3$, in which the central residuals alternate between saturated red and blue indicating the posterior samples are generally far from the ground truth values for these regions.
The second notable feature is that the structure of the residuals goes from high frequency alternating red/blue ``blobs'' for bright stars, to noticeably lower frequency fluctuations for faint stars.
Likely this is because of the broader posteriors for faint stars meaning the normalization has been smoothed accordingly.
Third, the residuals seem largely unstructured, with the most significant structures appearing in the central regions of stars $\#3$ and $\#5$.
This lack of structure is further evidence that the posterior samples are broadly ``surrounding'' the ground truth in the full high dimensional parameter space.
Fourth, the square border artifact is visible in the residuals from bright, down to faint stars, meaning that even when the square artifact is included in the ground truth, it shows up with some variability in the posterior samples.
This may be considered a positive for our model, as the border artifact is highly subdominant in the fitting; but ideally it would not exist at all.

We also checked the recovered fluxes to see if they matched the ground truth fluxes.
In all cases the recovered flux was indeed close to the true flux, with average deviations of approximately $1\%$.
However, for star $\#3$ we observed that the recovered posterior flux samples did not reflect the true uncertainty in the flux.
So while it was only off from the true value by $1.01\%$, this was an 8 sigma deviation given the distribution of the 32 flux estimates from the posterior.
This suggest, that the large deviations from the ground truth sample we saw in \Fig{inject} for star $\#3$ are from low variability in the posterior samples rather than a significant deviation from the gorund truth.
Likely all of our posterior distributions could benefit from broader sampling; which may be achieved by taking more than 32 posterior samples, training a broader prior and/or running the posterior sampling procedure with more variations in the nuisance parameters, running the posterior SDE exploration with finer step sizes and more corrector steps (or using algorithms beyond the convolved likelihood approximation discussed in \Sec{posterior}).

As a final test, we pass the ground truth PSF models and our posterior samples to PTED\footnote{Permutation Test using the Energy Distance, PTED can perform Bayesian coverage testing in scenarios where one does not have access to posterior density values. See the GitHub for more information: \url{https://github.com/ConnorStoneAstro/pted}.} to perform a coverage test~\citep{stone2026pted}.
The test returned a p-value of 0.002 meaning the null hypothesis (that the posterior has coverage over the ground truth\footnote{More specifically, the null hypothesis is that the ground truth sample was drawn from the posterior distribution and so is indistinguishable from any collection of randomly drawn posterior samples.}) is weakly rejected.
This is a clear suggestion that there are avenues for further improvement in our model.
A result that is not especially surprising given the limitations discussed throughout the text.
The test here suggests that the sampling procedure itself could use some further scrutinization.
This is consistent with what we discussed in the previous paragraph.
These are exciting avenues for future development.

\section{PSF Interpolation Test}\label{app:interpolate}
\begin{figure}
    \centering
    \includegraphics[width=0.45\linewidth]{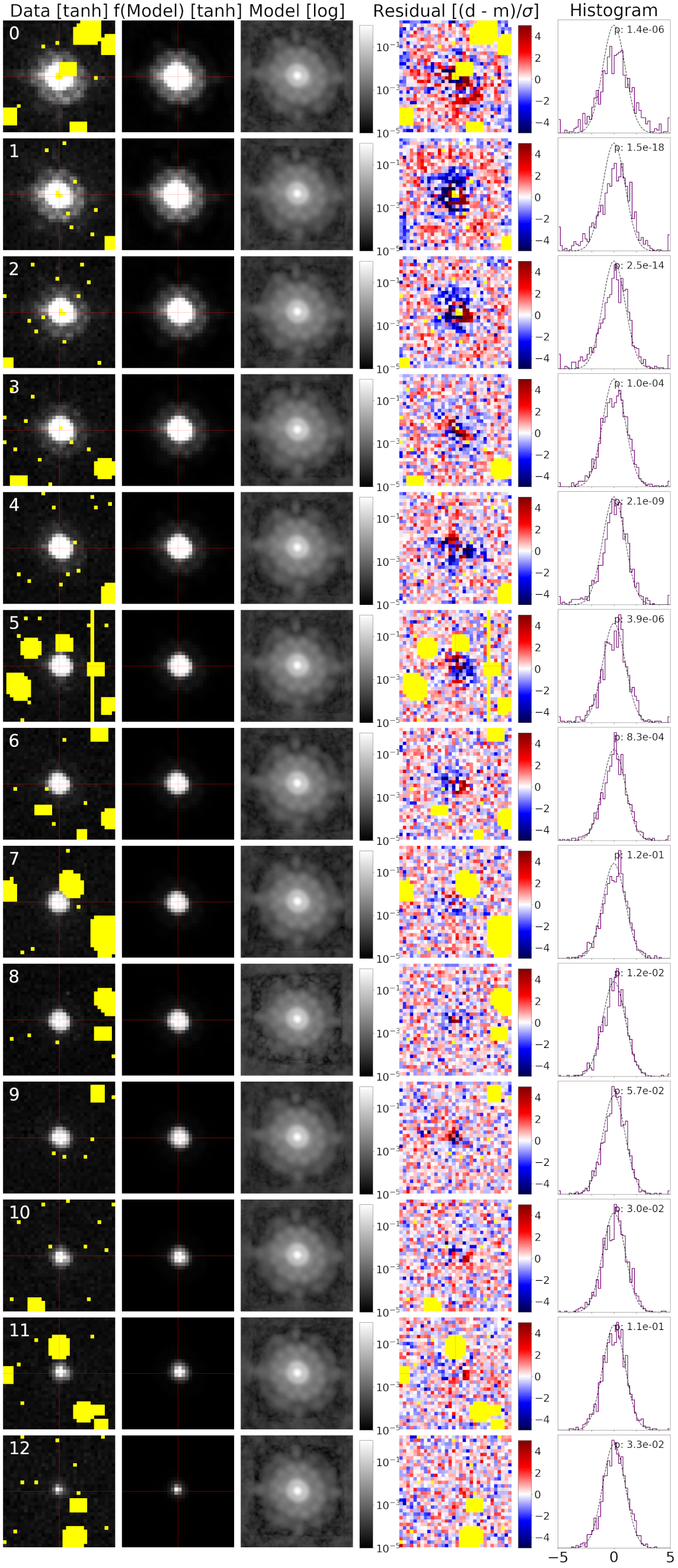}
    \includegraphics[width=0.45\linewidth]{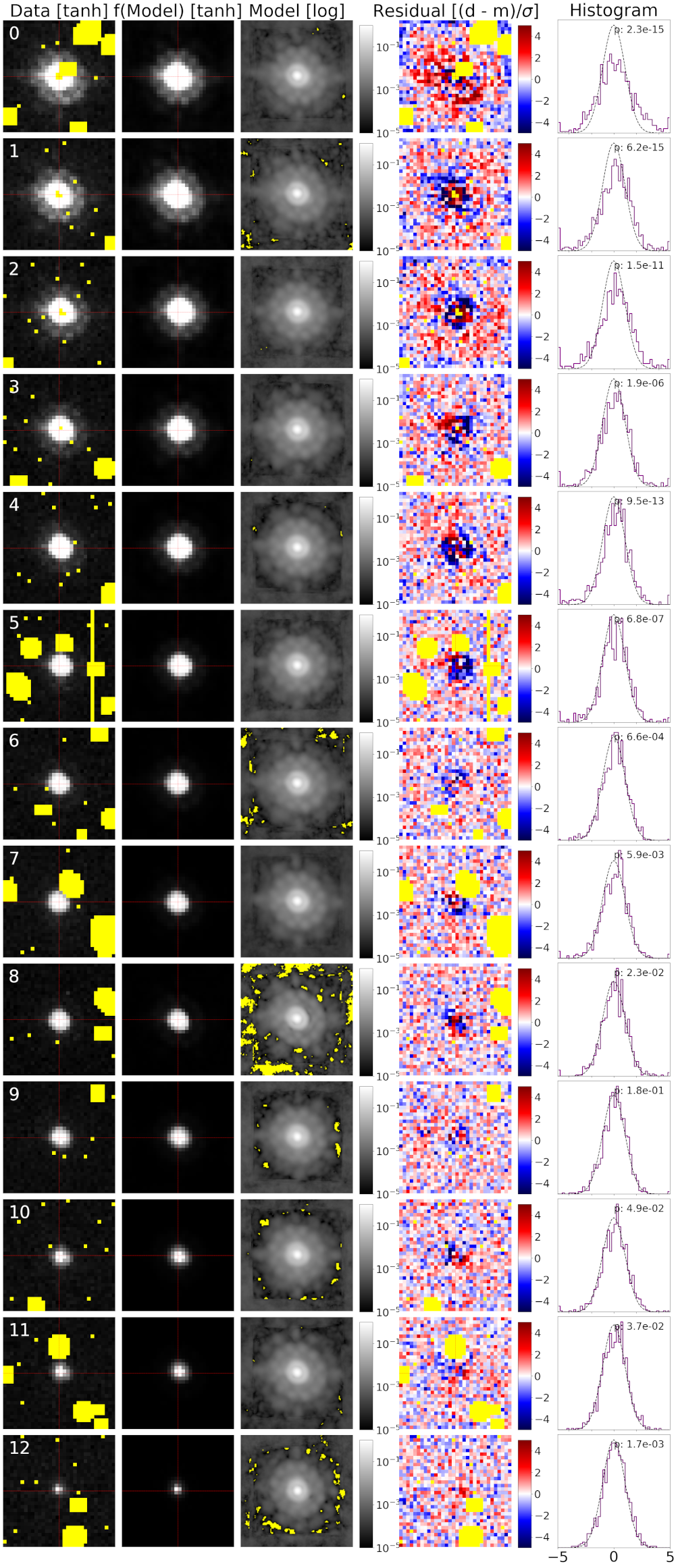}
    \caption{Cross-validation: interpolating the PSF to arbitrary positions. Format is identical to \Fig{astrophot}. This figure shows a leave-one-out test: each star was removed, and a new PSF was interpolated to its location using only the other stars on the frame. This interpolated model was then fit with an MLE to the held-out star. The resulting residuals are visually superior to all traditional methods in \Sec{old}. This validates our weighted-mean scheme (left) as a practical method for generating a PSF pseudo-posterior at any arbitrary position.}
    \label{fig:interpolate}
\end{figure}

In \Sec{interpolate}, we introduced a non-parametric, weighted-mean scheme to interpolate our PSF posterior to arbitrary detector locations. 
This appendix provides some validation for that pragmatic choice.
We compare two distinct pixel-wise interpolation schemes: 
\begin{enumerate}
    \item Weighted Mean: The method from \Sec{interpolate}, which weights each source's PSF model by its posterior variance and spatial proximity (using a broad 1024-pixel Gaussian kernel).
    \item Linear Fit: A scheme that fits a linear model in detector coordinates (a 2D plane) to the value of each pixel independently, weighting each source by its posterior variance.
\end{enumerate}
To test these methods, we perform a leave-one-out cross-validation. 
For each star in our sample, we remove it, generate an interpolated PSF at its location using only the remaining stars, and then fit this interpolated model to the held-out star's data (optimizing only for position, flux, and sky).

The results are presented in \Fig{interpolate}. 
As expected, the interpolated models perform worse than the direct posterior fits shown in \Fig{posterior}. 
As seen with the other tests, the brightest point sources are the most difficult to model sufficiently accurately, while the faint star residuals are consistent with noise even for the interpolated models.
However, they are visually superior to all three traditional methods shown in \Figto{astrophot}{psfex}, with residuals that are far more consistent with noise. 
This confirms that our framework, even when interpolated, provides a more accurate PSF model.

Comparing the two interpolation schemes, the weighted mean (left panel) produces slightly lower residuals and, critically, suffers from fewer ``catastrophic failures'' where the interpolation results in non-physical negative fluxes (seen as yellow pixels in the right panel). 
For this reason, we adopt the more robust weighted-mean scheme for our work, while acknowledging that a more rigorous, fully Bayesian interpolation model remains a promising avenue for future research.

\section{Flux Measurement Comparison}\label{app:flux}

\begin{figure}
    \centering
    \includegraphics[width=0.8\linewidth]{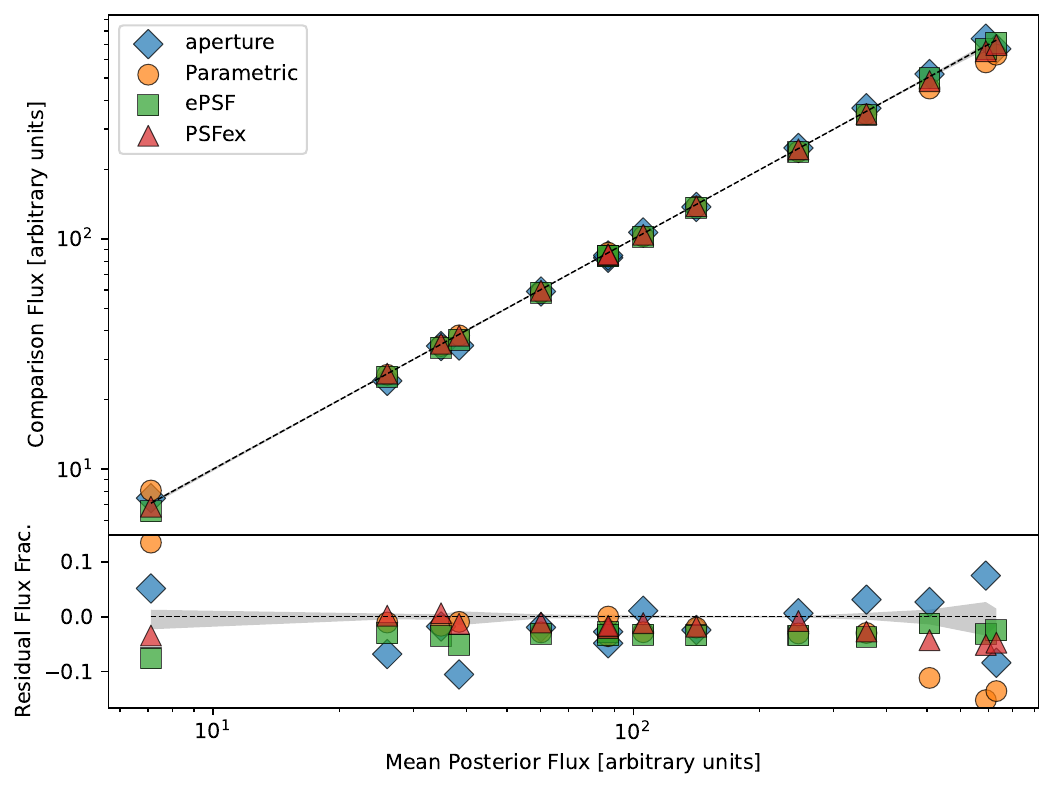}
    \caption{
    Biased photometry from misspecified priors. The posterior mean flux (x-axis) is compared against aperture, Moffat, ePSF template, and \texttt{PSFex} photometry (y-axis). A dashed line shows ideal agreement for reference. The lower subfigure shows the fractional residual flux ($\frac{F_{\rm estimate} - \bar{F}_{\rm posterior}}{\bar{F}_{\rm posterior}}$), including a grey band for the 16-84 percentile of our posterior flux values. The takeaway is that model misspecification leads to biased flux measurements: the traditional methods (Aperture, Moffat, ePSF, \texttt{PSFex}) show a consistent $\sim 5\%$ flux bias. Our posterior (grey band) provides an accurate, unbiased flux measurement with a principled uncertainty estimate which propagates morphological uncertainty.}
    \label{fig:flux}
\end{figure}

While flux estimation was not the primary goal of this analysis, it is instructive to compare the flux measurements derived from the various PSF fitting techniques. 
\Fig{flux} compares the single point-estimate fluxes from the traditional methods against the posterior mean and 16th-84th percentile range derived from our Bayesian posterior samples\footnote{Again with the outskirts trimmed for accuracy.}.
It is immediately clear that all traditional methods produce fluxes that are biased relative to our posterior estimates. 
The parametric Moffat model is the most severely affected; its significant model misspecification (see \Fig{astrophot}) translates directly into a large photometric bias.
The \texttt{PSFex} and ePSF template models perform more consistently as a function of flux, but both still exhibit a systematic $\sim 3\%$ bias, underestimating the flux for most point sources. 

Interestingly, there appears to be no systematic trend with flux, potentially indicating that our pixel flux limit (30,000 electrons) is successfully deselecting any non-linearities in our data.
We also include aperture flux values in the comparison by simply summing all pixels in the $32\times 32$ cutout (excluding masked pixels) after subtracting the initial sky estimate, though these should be considered inaccurate/imprecise as no correction was made for the masked pixels, which can be numerous, and no correction is made for light lost outside the cutout; both of these would bias the estimates low.
While most methods bias low, the aperture fluxes appear to trend to higher values than the other estimates for bright stars.
Investigating the cause of these biases, we discovered that the sky values for all other methods were biased high relative to our posterior samples.
Translating the sky level bias to a flux over the $32^2$ pixels resulted in a value that averaged about 2\% of the total flux for each point source.
This suggests that the discrepancy is largely due to sky level estimates alongside the point source flux.
It is possible the sky value was slightly ``compensating'' for larger residuals of bright pixels in the fit, at the expense of slightly offsetting all other faint pixels\footnote{A phenomenon known in least squares fitting, such as our likelihood model.}.
A complete explanation for this bias will require future investigations focused explicitly on precision photometry.
An avenue for future exploration could be in using a larger area than the cutout for inferring the sky level, though this introduces a number of complications beyond the scope of our current analysis (largely due to light from nearby and extended objects).

Overall, these results provide a quantitative warning: even a seemingly simple, summary statistic like total flux is highly sensitive to the accuracy of the underlying PSF model. 
This demonstrates that great care must be taken in a model's selection and validation to avoid propagating significant systematic biases into downstream scientific measurements.
Enticingly, this means there is likely opportunity for much higher precision in already available data, so long as models are sufficiently accurate.

\end{appendix}

\end{document}